\documentclass[aps,showpacs,amsmath,amssymb,twocolumn,nofootinbib]{revtex4-1}

\usepackage{subfigure}
\usepackage{graphicx}% Include figure files
\usepackage{dcolumn}% Align table columns on decimal point
\usepackage{bm}% bold math
\usepackage{color}
\usepackage[colorlinks=true,pdfstartview=FitV,linkcolor=blue,citecolor=blue,urlcolor=blue]{hyperref}
\usepackage[mathlines]{lineno}% Enable numbering of text and display math
\usepackage[dvipsnames]{xcolor}
\usepackage{amsmath}
\everymath{\displaystyle}
\begin{document}

\newcommand{\up}[1]{$^{#1}$}
\newcommand{\down}[1]{$_{#1}$}
\newcommand{\powero}[1]{\mbox{10$^{#1}$}}
\newcommand{\powert}[2]{\mbox{#2$\times$10$^{#1}$}}

\newcommand{\evm}{\mbox{\rm{eV\,$c^{-2}$}}}
\newcommand{\mevm}{\mbox{\rm{MeV\,$c^{-2}$}}}
\newcommand{\gevm}{\mbox{\rm{GeV\,$c^{-2}$}}}
\newcommand{\pgd}{\mbox{g$^{-1}$\,d$^{-1}$}}
\newcommand{\um}{\mbox{$\mu$m}}
\newcommand{\spix}{\mbox{$\sigma_{\rm pix}$}}
\newcommand{\pav}{\mbox{$\langle p \rangle$}}

\newcommand{\sige}{\mbox{$\bar{\sigma_e}$}}
\newcommand{\mass}{\mbox{$m_\chi$}}
\newcommand{\crystal}{\mbox{$f_\textnormal{c}(q,E_e)$}}
\newcommand{\electron}{\mbox{$\rm{e^-}$}}

\newcommand{\gc}[1]{\textcolor{magenta}{[#1]}}
\newcommand{\ms}[1]{\textcolor{blue}{[#1]}}

\newcommand{\beq}{\begin{equation}}
\newcommand{\eeq}{\end{equation}}
\newcommand{\beqs}{\begin{eqnarray}}
\newcommand{\eeqs}{\end{eqnarray}}
\newcommand{\MS}[1]{{\color{red}#1}}

\title{Degenerate Sub-keV Fermion Dark Matter from a Solution to the Hubble Tension}

\author{Gongjun Choi,$^{1}$}
\thanks{{\color{blue}gongjun.choi@gmail.com}}

\author{Motoo Suzuki,$^{1}$}
\thanks{{\color{blue}m0t@icrr.u-tokyo.ac.jp}}

\author{Tsutomu T. Yanagida,$^{1,2}$}
\thanks{{\color{blue}tsutomu.tyanagida@ipmu.jp}}

\affiliation{$^{1}$ Tsung-Dao Lee 
Institute, Shanghai Jiao Tong University, Shanghai 200240, China}
\affiliation{$^{2}$ Kavli IPMU (WPI), UTIAS, The University of Tokyo,
5-1-5 Kashiwanoha, Kashiwa, Chiba 277-8583, Japan}
\date{\today}

\begin{abstract}
We present a dark sector model addressing both the Hubble tension and the core-cusp problem. The model is based on a hidden Abelian gauge symmetry group with some chiral fermions required by the anomaly cancellation conditions, producing a candidate for the decaying fermion dark matter as a solution to the Hubble tension. Moreover, the sub-keV mass regime and the thermal history of the dark sector help the dark matter candidate resolve the core-cusp problem occurring in the standard $\Lambda$CDM cosmology.
\end{abstract}

\maketitle
\section{Introduction}  
The presence of dark matter (DM) becomes unquestionable fact thanks to various observational evidences. Nevertheless, its nature still remains unclear. So far, at least three physical features of DM are known, i.e., its stability, non-zero mass and very weak interaction with the ordinary matter.
Unfortunately, however, as for the three aspects of DM, none was examined clearly thus far. It still remains unanswered whether its stability is permanent or effective only for the time scale of the age of Universe today. Also, the uncertainty in a possible DM mass has not been narrowed down yet, ranging from $\mathcal{O}(10^{-22}){\rm eV}$ for the ultralight bosonic DM~\cite{Hu:2000ke,Hui:2016ltb} to $\mathcal{O}(10){\rm M}_{\odot}$ for primordial black holes
\footnote{See $e.g.$~\cite{Inomata:2017okj} for the possibility of primordial black holes as all dark matter.}
or  a macroscopic compact halo object (MACHO). Above all, lacking knowledge about non-gravitational interaction which DM does, we are still even unaware of if DM is, either of directly or indirectly, coupled to or totally decoupled from the SM sector.

In an effort to address aforementioned questions, we give a special attention to two arguments based on cosmological and astrophysical phenomena. The recently raised Hubble tension is one of them. This regards discrepancy reaching $\sim4\sigma$ level \cite{Verde:2019ivm} between a local measurement of the Hubble expansion rate ($H_{0}$) \cite{Riess:2016jrr,Riess:2018byc,Bonvin:2016crt,Birrer:2018vtm} and that inferred from the cosmic microwave background (CMB) observation \cite{Aghanim:2018eyx}. While unknown systematics may affect the discrepancy~\cite{Efstathiou:2013via,Freedman:2017yms,Rameez:2019wdt}, it could be a clue for a nature of DM.
In particular, a decaying DM (DDM) solution \cite{Vattis:2019efj}
\footnote{For an exemplary particle physics model for DDM solution to the Hubble tension, see \cite{Choi:2019jck,Feng:2003xh}.} 
to the tension amongst several resolutions proposes a possibility that the decay of DM with a lifetime $\sim\!\!35{\rm Gyrs}$ can relieve the tension.
\footnote{Other scenarios proposed to solve the tension include, for example, dark radiation~\cite{Riess:2016jrr,Bernal:2016gxb,Vagnozzi:2019ezj}, dark energy at early time~\cite{Karwal:2016vyq,Poulin:2018cxd,Alexander:2019rsc,Sakstein:2019fmf}, interacting dark energy~\cite{Yang:2018euj,DiValentino:2019ffd} etc.}
 The advanced starting of the dark energy dominated era due to DM decay enables the faster expansion of the Universe at late times ($z\!\simeq\!\mathcal{O}(0.1)$) and eventually the Hubble expansion rate at recombination era determined by CMB data can evolve to a value close to SHOES measurement of $H_{0}\simeq73-74{\rm Km/sec/Mpc}$ \cite{Riess:2018byc}. Inspired by this, we take a DM scenario with a finite lifetime greater than the current age of the Universe as far as the question about stability of DM is concerned.

On the other hand, whereas the correlation of structures on the large scale is in excellent accord with predictions from $\Lambda$CDM paradigm (cosmological constant + cold dark matter), discrepancies between results from simulations based on $\Lambda$CDM (e.g. cuspy halos \cite{Moore:1999gc}, too many subhalos \cite{Moore:1999nt,Kim:2017iwr}, dense cores \cite{Boylan_Kolchin_2011}) and the experimental observations on galactic scales may be hinting for a nature of DM distinct from CDM (for review, see \cite{Weinberg:2013aya}). Particularly, from disagreement between the cuspy DM density profile for galaxies predicted by $\Lambda$CDM simulations~\cite{Navarro:1996gj,Fukushige:1996nr,Ishiyama:2011af} and observations of rotation curves for low mass galaxies~\cite{Borriello:2000rv,Gilmore:2007fy,Oh:2008ww,deBlok:2009sp}, there arises the so-called cusp-core problem.
As one of resolutions to the problem, presence of degenerate fermion DM with sub-keV mass has been suggested in \cite{Destri:2012yn,Domcke:2014kla,Alexander:2016glq,Randall:2016bqw,Giraud:2018gxl,Savchenko:2019qnn}. The quantum pressure applied by a degenerate fermion gas can counterbalance the gravity to prevent gravitational collapse, yielding cored DM profiles in dwarf spheroidal (dSphs) galaxies. Modelling DM halo as a degenerate fermion gas with the use of the stellar velocity dispersion data of dSphs and the inferred core size of Fornax dwarf galaxy favor the fermion DM mass in sub-keV range. 

In this paper, we present a sub-keV fermion DDM scenario which addresses both the Hubble tension and the core-cusp problem. The model is based on a hidden Abelian gauge symmetry group and the fermion DDM is one of the chiral fermions required to cancel the chiral anomaly. The dark matter is non-thermally produced and sufficiently cooled down before virialization to follow the degenerate configuration. The Hubble tension is reconciled by the dark matter decaying into a massive dark photon and a light chiral fermion. We also comment about the constraint from the Lyman-$\alpha$ forest~\cite{Viel:2013apy,Irsic:2017ixq} and showed that our model is not inconsistent with its mass bound.

\section{A Model for the Dark Sector}
Let us begin with the basic framework of our model for the dark sector. As a minimal extension of the gauge sector in a full theory, we consider a hidden Abelian gauge symmetry $U(1)_{X}$ under which matters in the dark sector are charged while SM particles are neutral. 

As will be discussed below, the picture for DDM solution to the Hubble tension in \cite{Vattis:2019efj} assumes at least three particle candidates for (1) a decaying dark matter, (2) a radiation resulting from the decay and (3) a daughter dark matter resulting from the decay. Thus in the case where none of three is a SM particle, introduction of other fields in the dark sector than the $U(1)_{X}$ gauge field is necessary. Anomaly-free condition for $U(1)_{X}$ is an useful guide line for how the introduction is made. As a matter of fact, a simple set-up dubbed ``Number Theory Dark Matter" for the dark sector along this line of logic was already discussed in \cite{Nakayama:2011dj,Nakayama:2018yvj}. 

As for the total number of Weyl fermions ($N$) and their $U(1)_{X}$ charges ($Q_{i}$), anomaly-free conditions demand 
\beq
\sum_{i=1}^{N}Q_{i}^{3}=0\quad,\quad\sum_{i=1}^{N}Q_{i}=0\,,
\label{eq:anomalyfree}
\eeq
for cancellation of $U(1)_{X}^{3}$ anomaly and gravitational $U(1)_{X}\times[{\rm gravity}]^{2}$ anomaly. If the theory allows the vector-like fermions in the dark sector, values of their mass parameters are naturally close to a cut-off scale.\footnote{Of course, mass of a vector-like fermion is a free parameter and in principle it can be any value from zero to a UV-cut off of the theory. However, small mass regime is less appealing from the theoretical point of view unless there is any reasonable supporting explanation like restoration of certain symmetries in zero-limit.} This point makes the vector-like fermions become irrelevant to the low-energy physics once integrating out heavy degrees of freedom is done. Thus, we take an option to exclude presence of the vector-like fermions in the dark sector. DM being massive, this option automatically requires the breaking of $U(1)_{X}$. 

In order to have particle contents required for the DDM scenario, we realize that the minimum number of Weyl fermions in the dark sector satisfying Eq.~(\ref{eq:anomalyfree}) amounts to five.\footnote{Three Weyl fermions cannot satisfy Eq.~(\ref{eq:anomalyfree}) due to the Fermat's theorem. Both cases of two and four Weyl fermions satisfying Eq.~(\ref{eq:anomalyfree}) give rise to vector-like fermions, which is excluded in our option.} Apart from the gauge field and the five Weyl fields, we introduce two complex scalar fields for the spontaneous breaking of $U(1)_{X}$ and to make fermions massive. Among several possibilities for a set of $U(1)_{X}$ charges discussed in \cite{Nakayama:2011dj,Nakayama:2018yvj}, we consider the following charge assignment
\beq
\psi_{-9},\quad\psi_{-5},\quad\psi_{-1},\quad\psi_{7}\quad\psi_{8},\quad\Phi_{1},\quad\Phi_{6}\,,
\label{eq:U(1)charges}
\eeq
where the subscripts denote $U(1)_{X}$ charges of each field. The first five are Weyl fields and the remaining two are complex scalars. One can see that the anomaly free conditions in Eq.~(\ref{eq:anomalyfree}) are satisfied indeed by charges of Weyl fields. Writing the complex scalars as $\Phi_{1}=\phi_{1}e^{iA_{1}/V_{1}}/\sqrt{2}$ and $\Phi_{6}=\phi_{6}e^{iA_{6}/V_{6}}/\sqrt{2}$, we assume $V_{1}\equiv<\!\!\phi_{1}\!\!>$ is greater than $V_{6}\equiv<\!\!\phi_{6}\!\!>$ so that the breaking of $U(1)_{X}$ is induced dominantly by the condensation of $\Phi_1$ and hence the gauge boson mass is given by $m_A' \simeq gV_1$, where $g$ is the gauge coupling constant.\footnote{To have values of $V_{1}$ and $V_{6}$ discussed in Sec.~\ref{sec:aconcretescenario}, we suppress operators composed of both $\Phi_{1}$ and $\Phi_{6}$. To this end, as an exemplary way to do suppression as we desire, in an extra dimension picture we assume that the two complex scalars are localized in the different branes separated by a distance along an extra dimension which is large enough for suppressing effective interactions induced at 3+1 dimensional spacetime.}

The Yukawa coupling of the dark sector reads
\beqs
\mathcal{L}_{{\rm Yuk}}&=&y_{1}\Phi_{1}\psi_{-9}\psi_{8}\, +\,  y_{2}\Phi_{6}\psi_{-5}\psi_{-1}\cr\cr
&+&\,y_{3}\Phi_{6}^{\dagger}\psi_{-1}\psi_{7} \,+\, {\rm h.c.}\,,
\label{eq:yukawa}
\eeqs
where $y_{i}\,(i=1-3)$ are taken to be real parameters without loss of generality.
We diagonalize the mass matrix for $\psi_{-5}$, $\psi_{-1}$ and $\psi_{7}$ to obtain mass eigenstates. Together with $\psi_{-1}$, the following linear combination $\chi$ of $\psi_{-5}$ and $\psi_{7}$ 
\beq
\chi\equiv\left(\frac{y_{2}}{\sqrt{y_{2}^{2}+y_{3}^{2}}}\right)\psi_{-5}+\left(\frac{y_{3}}{\sqrt{y_{2}^{2}+y_{3}^{2}}}\right)\psi_{7}\,,
\eeq
forms a Dirac fermion $\Psi_{{\rm DM}}=(\psi_{-1},\chi^{*})^{T}$ of the mass $m_{\rm DM}\equiv\sqrt{y_{2}^{2}+y_{3}^{2}}\,V_{6}$. And also the other orthogonal combination to $\chi$ forms a massless Weyl field $\xi$
\beq
\xi\equiv\left(\frac{y_{3}}{\sqrt{y_{2}^{2}+y_{3}^{2}}}\right)\psi_{-5}-\left(\frac{y_{2}}{\sqrt{y_{2}^{2}+y_{3}^{2}}}\right)\psi_{7}\,.
\eeq

From here on, we assume $y_{2}$ and $y_{3}$ are of a similar order and $y_{2}\simeq y_{3}\equiv y_{*}$ for simplicity.  

\section{A concrete scenario}
\label{sec:aconcretescenario}
In this section, we discuss how each resolution to phenomenological problems can fit into our model for the dark sector presented above. To have a simple thermal history of the dark sector, we first assume that the mass ($\simeq y_{1}V_{1}$) of $\psi_{-9}$ and $\psi_8$ is greater than the inflaton mass $m_I$ so that we can neglect them in cosmological history of the Universe.%
\footnote{We also assume that the reheating temperature $T_{\rm RH}$ is much less than the inflaton mass.}

After the inflation ends, both the SM and the dark sec- tor are reheated by the inflaton decay. As a $U(1)_{X}$ singlet, the inflaton $\Phi_{I}$ couples to both $\Phi_{1}$ and $\Phi_{6}$ through the following renormalizable operators\footnote{In the extra dimension picture, taking the inflaton field $\Phi_{I}$ as the bulk field, we treat its coupling to $\Phi_{1}$ and $\Phi_{6}$ on an equal footing.}
\beq
\mathcal{L}_{\Phi}=b_{1}\Phi_{I}|\Phi_{1}|^{2}+b_{6}\Phi_{I}|\Phi_{6}|^{2}\,,
\eeq
where $b_{i}$(i=1,6) is a dimensionful coupling coefficient. Assuming quartic couplings $\lambda_{1}\simeq\lambda_{6}\simeq\mathcal{O}(1)$ for each of complex scalars $\Phi_{1}$ and $\Phi_{6}$, we realize that decay of the inflaton to two $\phi_{1}$s is kinematically suppressed because of $m_{I}\,<\,m_{1}\simeq\sqrt{\lambda_{1}}V_{1}$. Hence, for the dark sector, reheating is accomplished via inflaton decay to two $\phi_{6}$s. We define the decay rate to SM particles and the dark sector species ($\phi_{6}$) to be $\Gamma_{\rm{SM}}\equiv\Gamma(\Phi_{I}\rightarrow {\rm SM})$ and $\Gamma_{\rm{DS}}\equiv\Gamma(\Phi_{I}\rightarrow \phi_{6}+\phi_{6})$ respectively, and make an approximation $\Gamma_{\rm{tot}}=\Gamma_{\rm{SM}}+\Gamma_{{\rm DS}}\simeq\Gamma_{\rm{SM}}$. Unless a reheating temperature is so large as to be close to the Planck scale, $\phi_{6}$s form a dark thermal bath via scattering with one another among them. Thus, below we consider evolution of the dark thermal bath comprising purely $\phi_{6}$s.\footnote{In our work, we assume a negligibly small Higgs portal coupling between SM Higgs doublet and $\Phi_{6}$.} Notice that the purity of the thermal bath is guaranteed due to smallness of both $y_{*}$ and $U(1)_{X}$ gauge coupling ($g$) which will be shown later. 

The concrete thermal history which we will show below as a viable physical scenario is the following. After the reheating era, the dark sector begins with the temperature
\beqs
T_{{\rm DS}}(a_{{\rm RH}})\simeq 0.43\times\left(\frac{m_{{\rm DM}}}{1{\rm keV}}\right)^{-1/3}\times T_{{\rm RH}}\,,
\label{eq:TDSRH}
\eeqs
where $T_{\rm DS}(a)~(T_{\rm SM}(a))$ denotes the temperature of the dark sector (the SM sector) at the time of the scale factor $a$. $a_{{\rm RH}}$ is the scale factor at the reheating time, and $T_{{\rm RH}}\equiv T_{{\rm SM}}(a_{{\rm RH}})$ is assumed.%
\footnote{To get Eq.\,\eqref{eq:TDSRH}, the following relation is used,
\begin{equation}
Y_{\rm{DM}}\equiv
\frac{n_{\rm{DM}}}{s_{\rm{SM}}}\simeq4.07\times10^{-4}\times\left(\frac{m_{{\rm DM}}}{1{\rm keV}}\right)^{-1}\,,
\label{eq:DMA1}
\end{equation}
where $n_{\rm DM}$ is the number density of the dark matter, $s_{\rm SM}$ is the entropy density of the SM sector, and we used the values, $\Omega_{\rm{DM},0}=0.24$, $H_{0}=70\rm{km/sec/Mpc}$ \cite{Vattis:2019efj} and $s_{\rm{SM},0}\simeq2.945\times10^{-11}\rm{eV}^{3}$ (entropy density today). Using the approximation $2n_{\phi_{6}}=n_{\rm{DM}}$ at production time, one obtains
\begin{equation}
\rm{Br}\frac{T_{\rm{{\rm RH}}}}{m_{\Phi_{I}}}\simeq2.7\times10^{-4}\times\left(\frac{m_{{\rm DM}}}{1{\rm keV}}\right)^{-1}\,, 
\label{eq:DMA2}
\end{equation}
where Br is defined via $n_{6}={\rm Br}\times n_{I}\simeq{\rm Br}\times(\rho_{{\rm SM}}/m_{I})$ with $n_{6}$ and $n_{I}$ being the number density of $\phi_{6}$ and inflaton ($\Phi_{I}$) respectively.}
From here on, whenever we encounter $T_{{\rm RH}}$, we can convert it using $(a_{{\rm RH}}T_{{\rm RH}})=(a_{{\rm EW}}T_{{\rm EW}})\simeq10^{-13}{\rm GeV}$ with $a_{{\rm EW}}\simeq10^{-15}$ and $T_{{\rm SM}}(a_{{\rm EW}})\simeq100\,{\rm GeV}$. Afterwards, the temperature of the dark thermal bath continues to decrease as 
\beq
T_{{\rm DS}}(a)\simeq0.43\times\left(\frac{m_{{\rm DM}}}{1{\rm keV}}\right)^{-1/3}\!\!\times\frac{10^{-13}}{a}\,\,{\rm GeV}\,,
\eeq
and becomes lower than the mass of $\phi_{6}$. The comoving number density of $\phi_{6}$ is preserved until the time when the rate of $\phi_{6}$ decay to a DM pair ($\psi_{-1}$ and $\chi$) becomes comparable to the Hubble expansion rate, i.e., $\Gamma(\phi_{6}\!\rightarrow\!{\rm DM\!+\!DM})\simeq H$. Then, non-relativistic $\phi_{6}$s start to decay to produce free-streaming DMs. The free-streaming of DM is ensured owing to the small $y_{*}$. From $\Gamma(\phi_{6}\!\rightarrow\!{\rm DM\!+\!DM})\simeq H$, we infer the temperature of the SM thermal bath at this time
\beqs
T_{{\rm SM}}(a_{{\rm FS}})&&\,\,\simeq\,537\times g_{{\rm SM}}(a_{{\rm FS}})^{-1/4}\cr\cr
&&\times\left(\frac{V_{6}}{1{\rm GeV}}\right)^{-1}\left(\frac{m_{{\rm DM}}}{1{\rm keV}}\right)\sqrt{\frac{m_{6}}{1{\rm GeV}}}\,\,{\rm GeV}\,,
\label{eq:TSMFS}
\eeqs
where $a_{{\rm FS}}$ is the scale factor for the onset of the free-streaming of DM, $g_{{\rm SM}}(a)$ is the number of relativistic degrees of freedom in the SM sector evaluated at the scale factor $a$ and $m_{6}$ is the mass of $\phi_{6}$. Here we used $m_{{\rm DM}}\simeq y_{*}V_{6}$. Note that decay of $\phi_{6}$ and the onset of DM's free-streaming take place at the same time since $\phi_{6}$ is non-relativistic already at this time. For consistency, $T_{{\rm DS}}(a_{{\rm FS}})$ should be less than $m_{6}$, which constrains $V_{6}(\simeq\! m_{6}/\!\sqrt{\lambda_{6}}\!\simeq\! m_{6})$ for each $m_{{\rm DM}}$. In Fig \ref{fig:2}, the region above the green dashed line satisfies  $T_{{\rm DS}}(a_{{\rm FS}})\!<\!m_{6}$. Due to the entropy conservation, on the other hand, we can write down the temperature of the SM thermal bath as
\beq
T_{{\rm SM}}(a)=\left(\frac{106.75}{g_{{\rm SM}}(a)}\right)^{1/3}\frac{10^{-13}}{a}\,\,{\rm GeV}\,.
\label{eq:TSM}
\eeq
Equating Eq.~(\ref{eq:TSMFS}) with Eq.~(\ref{eq:TSM}) gives
\beqs
a_{\rm{{\rm FS}}}&&\,\,=\,8.8\times10^{-16}\times g_{{\rm SM}}(a_{\rm{{\rm FS}}})^{-1/12}\cr\cr
&&\times\left(\frac{V_{6}}{1{\rm GeV}}\right)\left(\frac{m_{{\rm DM}}}{1{\rm keV}}\right)^{-1}\left(\frac{m_{6}}{1{\rm GeV}}\right)^{-1/2}\,,
\label{eq:aFS}
\eeqs
which will be used later for computing the free-streaming length for DM.
After recombination, DM ($\Psi_{{\rm DM}}$) gradually decays to a massless radiation $\xi$ and a $U(1)_{X}$ gauge boson $A'_{\mu}$ serving as a daughter DM. On the other hand, due to a large redshift, DM gets into the motionless stage near recombination era, completely behaving as a matter. Furthermore, its ``would-be" temperature today without taking into account virialization due to gravity in the galaxy formation is sufficiently low to enable DM to be in degenerate configuration. 

In the following two subsections, we probe the parameter space of the model by applying the constraints on the lifetime of DDM and mass obtained in \cite{Vattis:2019efj,Destri:2012yn,Domcke:2014kla,Randall:2016bqw}. Eventually we show there exists a parameter space where the desired physics can be realized in the model.

\begin{figure*}[htp]
  \centering
  \hspace*{-5mm}
  \subfigure{\includegraphics[scale=0.349]{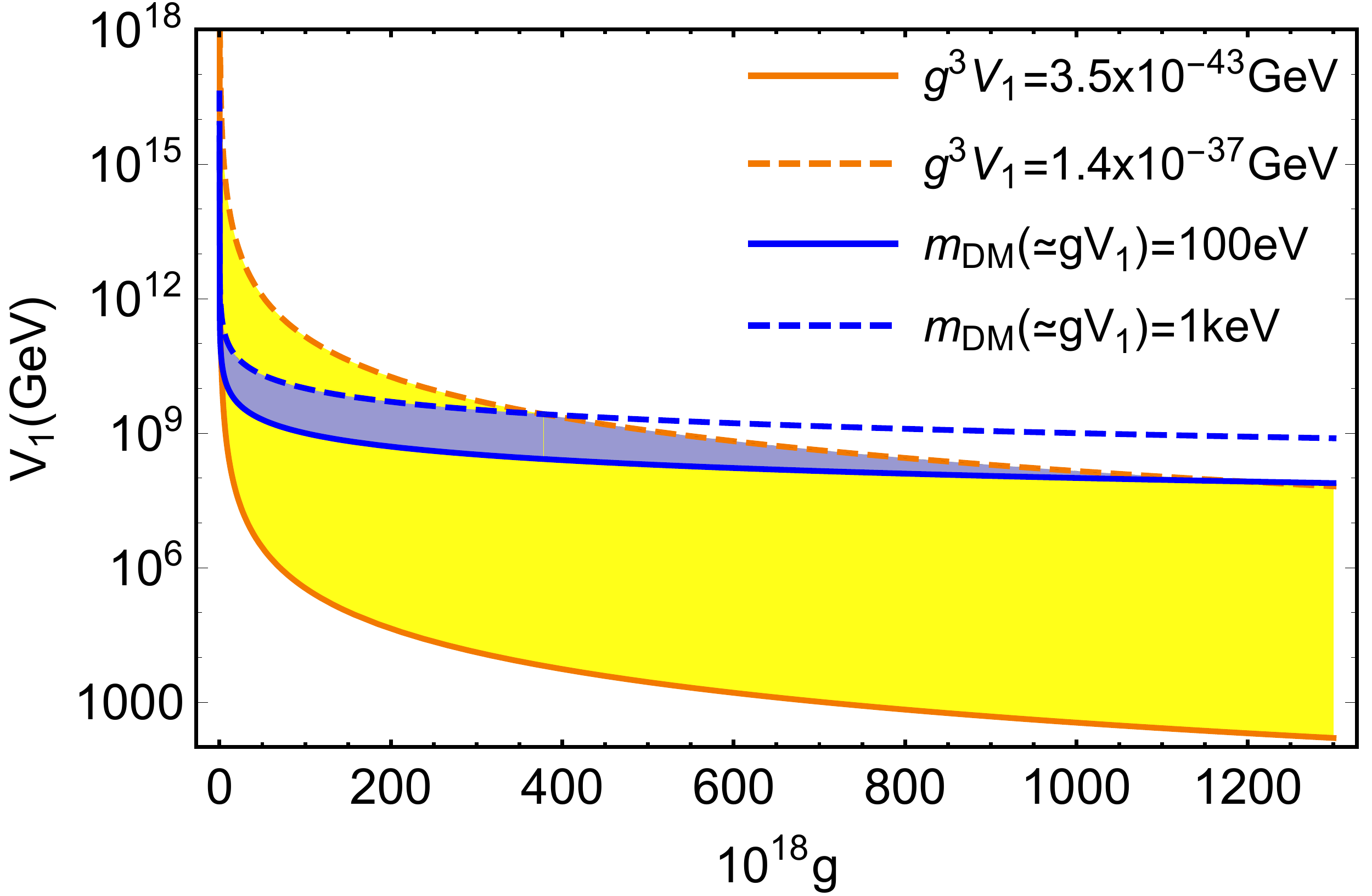}}\quad
  \subfigure{\includegraphics[scale=0.34]{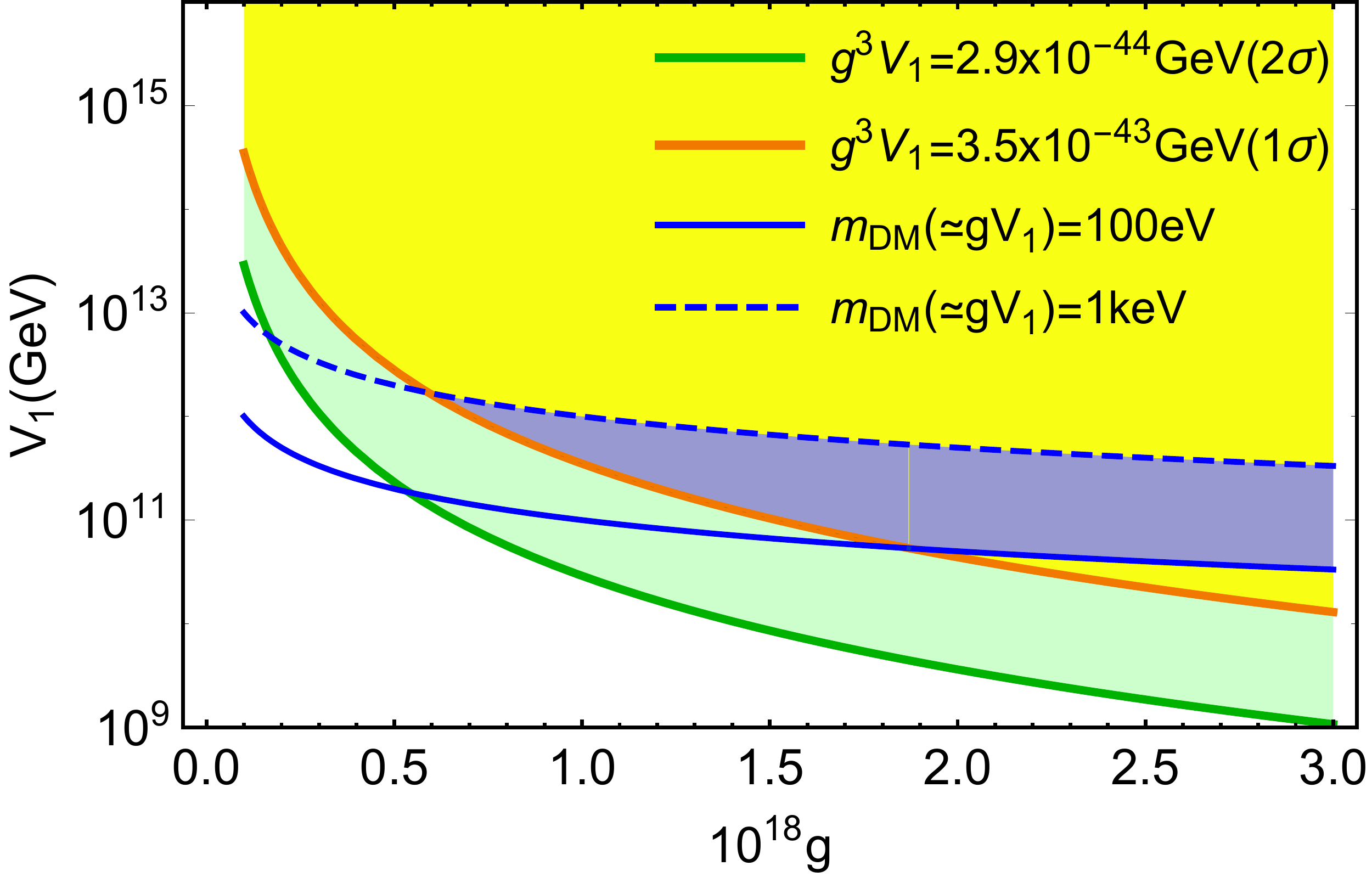}}
  \caption{Constraints on $(g,V_{1})$ space where $g$ is the gauge coupling of $U(1)_{X}$. The right panel is magnification of the left panel for the relatively smaller coupling regime. The yellow shaded region between the yellow solid and dashed line is mapping onto $(g,V_{1})$ space of constraints on the life of DDM and $\epsilon$ parameter in DDM solution to the Hubble tension \cite{Vattis:2019efj} at 1$\sigma$ level. (On the right panel, we showed the same mapping at 2$\sigma$ level with the green solid line and green shaded region.) The blue shaded region corresponds to $(g,V_{1})$ values that produce sub-keV decaying fermion DM as a solution to not only the Hubble tension, but the core-cusp problem, satisfying 1$\sigma$ level constraints.}
  \vspace*{-1.5mm}
\label{fig:1}
\end{figure*}

\subsection{Constraints from the Hubble Tension}
\label{sec:constraintHubble}
DDM solution to the Hubble tension is described by four parameters: lifetime of DDM, a fraction of a rest mass of DDM transferred to energy of the resulting radiation ($\epsilon$), DM abundance today $\Omega_{{\rm DM,0}}\equiv\rho_{{\rm DM,0}}/\rho_{{\rm cr,0}}$ and a reduced Hubble parameter $h\equiv H_{0}/{\rm Km/sec/Mpc}$ \cite{Vattis:2019efj}. Below we refer to constraints on these parameters reported in \cite{Vattis:2019efj} which were found by carrying out a Monte Carlo Markov Chain (MCMC) against the late universe measurements of the Hubble expansion rate for $0\!\lesssim\! z\!\lesssim\!2.4$. The underlying idea of the solution is to introduce a DDM whose decay leads to production of a massless and a massive particle with four momenta $p_{\mu}=(\epsilon m_{{\rm DM}}, \overrightarrow{p})$ and $p_{\mu}'=((1-\epsilon) m_{{\rm DM}},- \overrightarrow{p})$, respectively. Decrease in $\rho_{{\rm DM}}a^{3}$ after recombination due to decay of DDM is compensated by increase in $\rho_{\Lambda}$, which induces the earlier transition from the matter-dominated era to the dark energy dominated era. In the end, $H(a)$ evolves to a larger $H_{0}$ value than that resulting from CMB based on $\Lambda$CDM. As phenomenological consequences, enhancement of the late integrated Sachs-Wolfe (LISW) effect for low $\ell$ regime ($\ell\sim10$) can be induced regarding CMB temperature anisotropy power spectrum.  \cite{Poulin:2016nat,Xiao:2019ccl}. Also, reduction of $f\sigma_{8}$ in comparison with that from $\Lambda$CDM and suppression of the kinetic Sunyaev-Zel'dovich (kSZ) power spectrum were further discussed in  \cite{Xiao:2019ccl} as impacts that DDM can make. (Here $f$ is the matter growth rate, defined to be $f\equiv d\ln\delta_{m}/d\ln a$)

In our model, we identify the Dirac fermion $\Psi_{{\rm DM}}=(\psi_{-1},\chi^{*})^{T}$ generated from the decay of $\phi_{6}$ with DDM, the gauge boson of $U(1)_{X}$ with the resulting masssive particle, and the massless Weyl fermion $\xi$ with the resulting massless radiation. The dispersion relation of the resulting massive particle gives
\beq
\frac{m_{A^{'}}}{m_{{\rm DM}}}=\sqrt{1-2\epsilon}\,.
\label{eq:massratio}
\eeq

From the coupling of the Dirac fermion DM to the gauge boson, the decay rate of DDM can be read as 
\beqs
\Gamma_{{\rm DM}\rightarrow A^{'}\xi}&&=\frac{9}{4\pi}\epsilon g^{2}m_{A'}\left[\left(\frac{m_{{\rm DM}}}{m_{A'}}\right)^{3}\!\!-2\frac{m_{A'}}{m_{{\rm DM}}}+\frac{m_{{\rm DM}}}{m_{A'}}\right]\cr\cr
&&=\frac{9g^{2}m_{A'}}{2\pi}\times\frac{\epsilon^{2}(3-4\epsilon)}{(1-2\epsilon)^{3/2}}
\,.
\label{eq:GammaDM}
\eeqs
Applying $1\sigma$ level constraints $1.376\times10^{-43}{\rm GeV}\lesssim\Gamma_{{\rm DM}\rightarrow A^{'}\xi}\lesssim1.04\times10^{-42}{\rm GeV}$~%
\footnote{The constraint on DDM lifetime discussed in Refs.~\cite{Enqvist:2015ara,Enqvist:2019tsa} is not applied here because DDM only decaying into dark radiations is assumed there.}
and $0.0013\lesssim\epsilon\lesssim0.229$ \cite{Vattis:2019efj} to Eq.~(\ref{eq:GammaDM}), we obtain the constraints on $(g,V_{1})$ which is shown as the yellow shaded region in Fig \ref{fig:1}. The right panel is the magnification of the left panel for the smaller coupling regime where not only 1$\sigma$ but also 2$\sigma$ level is shown as green shaded region (the region above the green solid line).%
\footnote{The lifetime of $A'$ can be longer than that of $\Psi_{\rm DM}$ depending on $y_2$ and $y_3$.}

\subsection{Constraints from the Core-Cusp Problem}
\label{sec:constraintcorecusp}
Because of DM's fermionic nature, provided that a thermal history of DM can allow for its degenerate configuration near the time of the structure formation, then our DM candidate may form cored density profiles in Milky way's dSphs satellites by balancing the attractive gravitational force with the repulsive Fermi quantum pressure. To see whether our DM can go through sufficient redshift for achieving the degenerate configuration, we need to estimate DM's ``would-be" temperature today, $\tilde{T}_{{\rm DM,0}}$, in the absence of structure formation and compare it to a degeneracy temperature, $T_{{\rm DEG}}$, for dSphs to check if $\tilde{T}_{{\rm DM,0}}$ is less than $T_{{\rm DEG}}$ of dSphs \cite{Domcke:2014kla}. Starting with $T_{{\rm DM}}(a_{{\rm FS}})\simeq m_{6}/2$, we obtain $T_{{\rm DM}}(a_{{\rm NR}})\!\simeq\! (m_{6}a_{{\rm FS}})/(2a_{{\rm NR}})\simeq m_{{\rm DM}}$ where $a_{{\rm NR}}$ is the scale factor when DM becomes non-relativistic. Combined with Eq.~(\ref{eq:aFS}), this yields
\beqs
a_{NR}&&=4.4\times10^{-10}\times g_{{\rm SM}}(a_{{\rm FS}})^{-1/12}\cr\cr
&&\times\left(\frac{V_{6}}{1{\rm GeV}}\right)\left(\frac{m_{{\rm DM}}}{1{\rm KeV}}\right)^{-2}\left(\frac{m_{6}}{1{\rm GeV}}\right)^{1/2}\,.
\label{eq:aNR}
\eeqs
Since $a=a_{{\rm NR}}$, $T_{{\rm DM}}$ scales as $\sim a^{-2}$, implying that
\beq
\tilde{T}_{{\rm DM,0}}\simeq\frac{m_{6}a_{{\rm FS}}}{2a_{NR}}\left(\frac{a_{NR}}{a_{0}}\right)^{2}\,,
\label{eq:wouldbeTDM}
\eeq
where $a_{0}=1$ is the scale factor today. Below we will see values of $V_{1}$ obtained from the free-streaming length constraint lie in $V_{6}\simeq\mathcal{O}(10)-\mathcal{O}(100){\rm GeV}$. Using Eq.~(\ref{eq:aNR}) and Eq.~(\ref{eq:wouldbeTDM}) with $V_{6}\simeq\mathcal{O}(10)-\mathcal{O}(100){\rm GeV}$, we estimate ``would-be" temperature today of sub-keV DM to obtain $\mathcal{O}(10^{-14})-\mathcal{O}(10^{-11}){\rm eV}$. This corresponds to $\mathcal{O}(10^{-10}){\rm K}-\mathcal{O}(10^{-6}){\rm K}$ which is lower than the degeneracy temperature for dSphs $T_{{\rm DEG}}\simeq\mathcal{O}(10^{-4}){\rm K}-\mathcal{O}(10^{-3}){\rm K}$ \cite{Domcke:2014kla}. Hence, it is confirmed that our DM is capable of achieving the degenerate configuration today when the heat-up due to the virialization is neglected.  

As such, the mass of our DM is subject to Tremaine-Gunn bound \cite{Tremaine:1979we} which arises due to presence of maximum phase space density of the degenerate Fermi gas in a galaxy. When applied to dSphs, for a generic fermion DM, the logic underlying the Tremaine-Gunn bound gives $m_{{\rm DM,min}}\sim70-300\,{\rm eV}$ as a lower bound on mass \cite{Boyarsky:2008ju,Randall:2016bqw}. Also, recently, a fitting analysis for the stellar kinematics of the relatively smaller dSphs (Leo II, Willman I, Segue I) set a conservative mass lower bound $\sim100\,{\rm eV}$ ~\cite{DiPaolo:2017geq}.%
\footnote{These small size galaxies are particularly important for studying possibility of a degenerate fermion DM and its mass since increase in DM temperature due to virialization is minimized. For large size galaxies, fermion DM would be present there in non-degenerate configurations, which makes inferring fermion DM mass from the stellar kinematics of the large size galaxies more difficult.} In the light of these results, we focus on the fermion DM mass greater than $\sim100\,{\rm eV}$ in our model.  Combined with constraint on $\epsilon$ parameter in DDM solution to the Hubble tension, Eq.~(\ref{eq:massratio}) tells us that $m_{A'}$ is at least $70\%$ of $m_{{\rm DM}}$. Thus, we may approximate $m_{{\rm DM}}\simeq m_{A'}=gV_{1}$ and apply $m_{{\rm DM}}>100{\rm eV}$ to $(g,V_{1})$ plane to improve constraints obtained in Sec~\ref{sec:constraintHubble}. In Fig \ref{fig:1}, $m_{{\rm DM}}=100{\rm eV}$ is represented as the blue solid line. Thereby for the right panel the allowed parameter space at 1$\sigma$ (2$\sigma$) level reduces to the sub-area of the yellow (green) colored region which intersects with the region above the blue solid line. 

Next, in order for the degenerate fermion DM to serve as a solution to the core-cusp problem, it cannot have too large a mass because the core-size of the fermion DM density profile tends to decrease when $m_{{\rm DM}}$ increases. In this context, there have been efforts to constrain fermion DM mass by using the kinematic of dSphs. In \cite{Domcke:2014kla}, the use of fermion DM density profile obtained by solving Lane-Emden equation achieves a good fit to the stellar velocity dispersion of eight classical dwarf galaxies especially for the DM mass regime $100\!-\!200\,{\rm eV}$.\footnote{The Lane-Emden equation results from combining the continuity equation and the hydrostatic equilibrium equation with the Fermi pressure.} In addition, in \cite{Randall:2016bqw}, the use of the core-size of Fornax dSphs gives $70\,{\rm eV}\lesssim m_{{\rm DM}}\lesssim400\,{\rm eV}$ under the assumption of quasi-degenerate Fermi gas as DM. In contrast, relatively large fermion DM mass near $1\!-\!2\,{\rm keV}$ was pointed out in \cite{Destri:2012yn} by requiring the mass of dSph Willman I to be greater than the minimum halo mass made up of the degenerate fermion DM. Given these arguments for $m_{{\rm DM}}$, in this work we concentrate on sub-keV fermion DM mass regime, i.e. $m_{{\rm DM}}\!<\!1{\rm keV}$. Similarly to the previous section, we apply $m_{{\rm DM}}\lesssim1{\rm keV}$ to $(g,V_{1})$ plane to get the blue dashed line and area below it in Fig \ref{fig:1}. As the final intersection of several constraints at 1$\sigma$ level discussed so far, the blue shaded region in Fig \ref{fig:1} is obtained, which allows our model to produce {\it a degenerate decaying fermion DM} resolving the Hubble tension and the core-cusp problem.\footnote{In the right panel, the intersection coming from constraints at 2$\sigma$ level is the subspace between the blue solid and dashed lines which overlaps the green shaded region above the green solid line.}

Intriguingly, we observe that the blue shaded region (1$\sigma$) in Fig \ref{fig:1} can cover $V_{1}$ value up to $10^{12}{\rm GeV}$ for the gauge coupling as small as $g\simeq10^{-18}$. Besides, when 2$\sigma$ level constraints on the $(g,V_{1})$ plane is considered, even $V_{1}$ value as large as $10^{13}{\rm GeV}$ is allowed. As we mentioned above, $V_{1}$ is assumed to be larger than the inflaton mass $m_{I}$ in the model.  This implies that the model can be consistent with even the high scale inflation models such as chaotic inflation \cite{Nakayama:2013nya} or topological inflation \cite{Harigaya:2012hn} where the inflaton mass is $m_{I}\simeq10^{13}{\rm GeV}$.

\subsection{Constraints from the Free-Streaming Length}
\begin{figure}[t]
\centering
\hspace*{-5mm}
\includegraphics[width=0.49\textwidth]{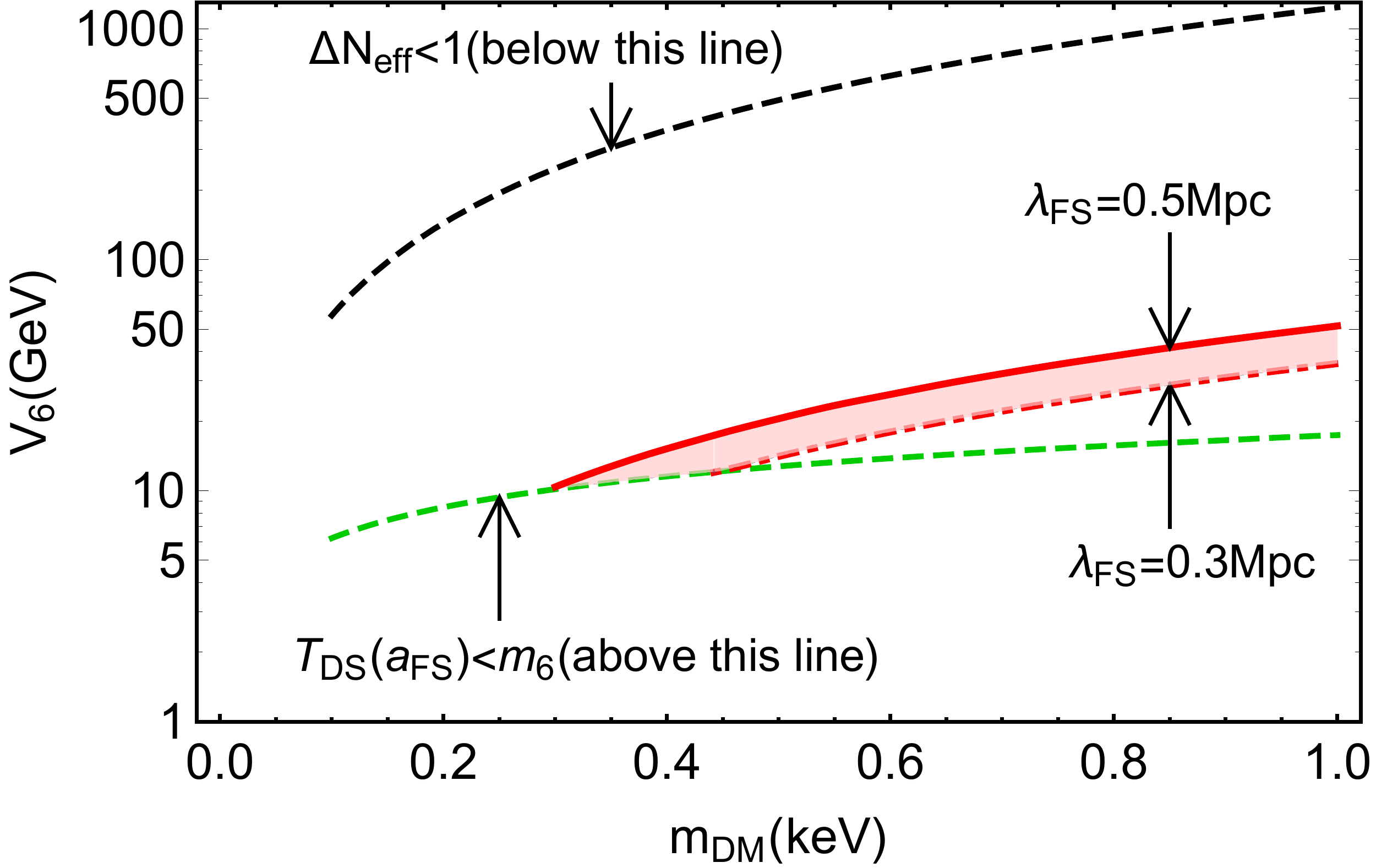}
\caption{$(m_{{\rm DM}},V_{6})$ plane which is constrained by (1) $T_{{\rm DS}}(a_{{\rm FS}})\!\!<\!\!m_{6}$ (region above the green dashed line), (2) $\Delta N_{{\rm eff}}^{{\rm BBN}}\!\!\lesssim\!\!1$ (region below the black dashed line), (3) $0.3{\rm Mpc}\!\!<\!\!\lambda_{{\rm FS}}\!<\!0.5{\rm Mpc}$ (region between the solid and dotdashed red lines). The eventual intersection of these lies in the red colored shaded region.}
\vspace*{-1.5mm}
\label{fig:2}
\end{figure}

Here we discuss constraints on Yukawa coupling $y_{*}$ as well as $V_{6}(\!\simeq\! m_{6})$. As a decay product of a non-relativistic $\phi_{6}$ which was never in a thermal equilibrium with other species, our Dirac fermion DM ($\Psi_{{\rm DM}}$) is expected to follow delta-function like distribution in its momentum space, centered on $|\vec{p}|\!=\!m_{6}/2$ at $a\!=\!a_{{\rm FS}}$. From then on, DM becomes non-relativistic around the time $a=a_{{\rm NR}}$ given in Eq.~(\ref{eq:aNR}). Bearing in mind $a_{{\rm BBN}}\simeq10^{-10}\!\!-\!\!10^{-9}$, we realize that for sub-keV DM, $V_{6}(\simeq m_{6})$ around GeV scale or greater than that makes DM still relativistic at BBN era. DM may contribute to energy budget of the universe at BBN era as a radiation. 
Therefore, we apply $\Delta N_{{\rm eff}}^{{\rm BBN}}\lesssim1$ \cite{Mangano_2011} to DM energy density at BBN era to constrain $V_{6}$. 
Using Eq.~(\ref{eq:DMA1}) and Eq.~(\ref{eq:aFS}) with
\beq
\rho_{{\rm DM}}(a_{{\rm BBN}})=\sqrt{m_{{\rm DM}}^{2}+\left(\frac{m_{6}a_{{\rm FS}}}{2a_{{\rm BBN}}}\right)^{2}}\,Y_{{\rm DM}}\,s_{{\rm SM}}(a_{{\rm BBN}})\,,
\label{eq:rhoDMBBN}
\eeq
yields the parameter space in $(m_{{\rm DM}},V_{6})$ plane satisfying the condition $\Delta N_{{\rm eff}}^{{\rm BBN}}\lesssim1$. In Fig \ref{fig:2}, this corresponds to the region below the black dashed line.

Given constraint on a value of $V_{6}$ around $\mathcal{O}(10)-\mathcal{O}(100){\rm GeV}$ (the region between the black and green dashed lines in Fig \ref{fig:2}), now it becomes clear that sub-keV DM is produced from $\phi_{6}$-decay as a relativistic particle. So the next question that should be further asked is whether DM in the model can be consistent with Lyman-$\alpha$ forest data or not. Should the free-streaming length ($\lambda_{{\rm FS}}$) of DM is too large, the structure formation at the scale below $\lambda_{{\rm FS}}$ would have been erased, likely to cause a disagreement with non-vanishing matter power spectrum at the large scales. On the contrary, if $\lambda_{{\rm FS}}$ is too short (i.e. $\lambda_{{\rm FS}}\!<\!0.1{\rm Mpc}$), DM model tends to produce too many satellite galaxies of Milky and Andromeda as opposed to the observed number. Therefore, we require $0.3{\rm Mpc}\!<\!\lambda_{{\rm FS}}\!<\!0.5{\rm Mpc}$ to further constrain $V_{6}$ \cite{Bringmann:2007ft,Domcke:2014kla,Cembranos:2005us,Colin:2000dn}. 

The free-streaming length is given by
\beqs
\lambda_{\rm{{\rm FS}}}&&=\int_{t_{\rm{{\rm FS}}}}^{t_{0}}\frac{v(t)}{a(t)}dt\cr\cr
&&\simeq\!\!\int_{a_{\rm{{\rm FS}}}}^{1}\!\!\frac{(a_{\rm{{\rm FS}}}m_{6}/H_{0})da}{\sqrt{\Omega_{{\rm rad},0}\!+\!a\Omega_{{\rm m},0}}\!\!\sqrt{(2am_{{\rm DM}})^{2}\!+\!(a_{\rm{{\rm FS}}}m_{6})^{2}}}\,,
\label{eq:FS}
\eeqs
where $v(t)$ is the DM velocity, and $\Omega_{\rm rad,0},~\Omega_{\rm m,0}$ denote the radiation and matter density parameter, respectively.
The dependence of $V_{6}$ appears through $a_{\rm{{\rm FS}}}$ in Eq.~(\ref{eq:aFS}). We show the result of mapping $0.3{\rm Mpc}\!<\!\lambda_{{\rm FS}}\!<\!0.5{\rm Mpc}$ to $(m_{{\rm DM}},V_{6})$ plane in Fig \ref{fig:2} as the space between the two red solid and dotdashed lines. The two red lines correspond to each of $\lambda_{{\rm FS}}=0.5$Mpc and 0.3Mpc. The eventual intersection in $(m_{{\rm DM}},V_{6})$ plane reflecting four constraints discussed so far is shown as the red shaded region in Fig \ref{fig:2}. We found that for the mass regime as low as $100-300{\rm eV}$, the non-thermally produced DM in our model travels the larger $\lambda_{{\rm FS}}$ than 0.5Mpc while there exists a consistent parameter space for $300-1000{\rm eV}$. The strength of the Yukawa $y_{*}$ associated with this $V_{6}$ regime reads $10^{-8}\!\!-\!10^{-7}$. This may appear to be in tension with \cite{Domcke:2014kla,Randall:2016bqw}. However, note that to resolve the core-cusp problem with $100{\rm eV}\!\lesssim\! m_{{\rm DM}}\!\lesssim\!400{\rm eV}$, either of a fully or quasi degenerate fermion DM configuration was assumed in \cite{Domcke:2014kla,Randall:2016bqw} in the absence of considering a baryonic effect in classical dSphs. As was pointed out in \cite{Randall:2016bqw}, when additional effects such as baryonic feedback or non-trivial DM momentum distribution skewed to lower energies are taken into account together with Fermi repulsion, the upper bound of $m_{{\rm DM}}$ might be alleviated.
 Additionally, in our work, we expect actual momentum distribution of our DM on production to have a broader width due to non-zero velocity dispersion of DM when produced and a center smaller than $m_{6}/2$ although we approximated it as a delta-function like peak with the assumption of instantaneous decay of $\phi_{6}$. This consideration may help some part of $100{\rm eV}\!\lesssim\! m_{{\rm DM}}\!\lesssim\!300{\rm eV}$ in Fig \ref{fig:2} produce the smaller $\lambda_{{\rm FS}}$ than we obtained in the present analysis.

\section{Discussion}
In this letter, we present a particle physics model which describes a completely isolated dark sector lacking any way to communicate with the SM sector. The dark sector enjoys an Abelian gauge symmetry $U(1)_{X}$ which meets the anomaly free conditions solely due to fields in the dark sector. We assume $U(1)_{X}$ without the kinetic mixing with $U(1)_{{\rm EM}}$ in the SM. This set-up trivially does not affect any known experimental result in the SM. With the purpose to keep the model minimal and consistent with the decaying DM solution to the Hubble tension, we introduced two darkly charged complex scalars and five darkly charged Weyl fermions along with the $U(1)_{X}$ gauge boson. Starting from an inflation model with inflaton mass smaller than $U(1)_{X}$ breaking scale, we showed that one linear combination of two of five Weyl fermions with sub-keV mass can play a role of DM which starts to decay after recombination with the lifetime $\sim35{\rm Gyr}$. Within the parameter space of the model we found, the redshift that DM candidate experiences since its production from decay of a complex scalar makes ``would-be" temperature of DM today without virialization low enough to form a cored density profile for Milky way's dwarf spheroidal satellites. The main conclusion in this paper is based on the DM production mechanism. Although we restrict our discussion to the specific model, the similar conclusions regarding free-streaming length of DM will be obtained as long as DM production mechanism presented in this work is used.

As such, DM candidate in our model is expected to serve as a solution to both the Hubble tension and the core-cusp problem. Originated from the dark sector lacking coupling to the SM sector, our DM candidate interacts with the SM particles only via gravitational interaction. We found that for resolving the Hubble tension, DM needs to decay via the gauge interaction with the coupling as tiny as $\sim\!\!10^{-18}\!\!-\!\!10^{-15}$ for having the required long life time. Interestingly, it turns out in the model that the strength of $U(1)_{X}$ gauge coupling is also linked to the a scale of consistent inflation models. Particularly, inflation models with inflaton mass as large as $10^{13}{\rm GeV}$ require the gauge coupling as small as $10^{-18}$. For sub-keV DM mass regime required to solve the core cusp problem, this coupling strength is still larger than the strength of gravitational interaction that DM does, i.e., $\sim\! m_{{\rm DM}}/M_{P}\!\simeq\!10^{-24}$. As for the sub-keV DM mass regime, we found that DM mass allowed in the model is in a slight tension with what is required for a degenerate fermion DM solution to the core-cusp problem, which we believe can be alleviated by a more detailed analysis about the phase-space distribution of DM and the use of unknown baryonic effects on dSphs galaxy formation.%
\footnote{Another possibility to alleviate the tension is to produce the fermion DM non-thermally by the inflaton decay. In this case, even the dark matter mass as low as $m_{\rm DM}\simeq200$eV can achieve a short enough free-streaming length~\cite{Domcke:2014kla}. For the same purpose, we also refer the readers to Ref.~\cite{Choi:2020nan} where we study more diverse non-thermal DM production mechanisms in another model of a fermion DM.}   \\

% ==================================================================

\begin{acknowledgments}
T. T. Y. is supported in part by the China Grant for Talent Scientific Start-Up Project and the JSPS Grant-in-Aid for Scientific Research No. 16H02176, No. 17H02878, and No. 19H05810 and by World Premier International Research Center Initiative (WPI Initiative), MEXT, Japan. 

\end{acknowledgments}

% ================================================================

\bibliography{main}

\begin{thebibliography}{59}%
\makeatletter
\providecommand \@ifxundefined [1]{%
 \@ifx{#1\undefined}
}%
\providecommand \@ifnum [1]{%
 \ifnum #1\expandafter \@firstoftwo
 \else \expandafter \@secondoftwo
 \fi
}%
\providecommand \@ifx [1]{%
 \ifx #1\expandafter \@firstoftwo
 \else \expandafter \@secondoftwo
 \fi
}%
\providecommand \natexlab [1]{#1}%
\providecommand \enquote  [1]{``#1''}%
\providecommand \bibnamefont  [1]{#1}%
\providecommand \bibfnamefont [1]{#1}%
\providecommand \citenamefont [1]{#1}%
\providecommand \href@noop [0]{\@secondoftwo}%
\providecommand \href [0]{\begingroup \@sanitize@url \@href}%
\providecommand \@href[1]{\@@startlink{#1}\@@href}%
\providecommand \@@href[1]{\endgroup#1\@@endlink}%
\providecommand \@sanitize@url [0]{\catcode `\\12\catcode `\$12\catcode
  `\&12\catcode `\#12\catcode `\^12\catcode `\_12\catcode `\%12\relax}%
\providecommand \@@startlink[1]{}%
\providecommand \@@endlink[0]{}%
\providecommand \url  [0]{\begingroup\@sanitize@url \@url }%
\providecommand \@url [1]{\endgroup\@href {#1}{\urlprefix }}%
\providecommand \urlprefix  [0]{URL }%
\providecommand \Eprint [0]{\href }%
\providecommand \doibase [0]{http://dx.doi.org/}%
\providecommand \selectlanguage [0]{\@gobble}%
\providecommand \bibinfo  [0]{\@secondoftwo}%
\providecommand \bibfield  [0]{\@secondoftwo}%
\providecommand \translation [1]{[#1]}%
\providecommand \BibitemOpen [0]{}%
\providecommand \bibitemStop [0]{}%
\providecommand \bibitemNoStop [0]{.\EOS\space}%
\providecommand \EOS [0]{\spacefactor3000\relax}%
\providecommand \BibitemShut  [1]{\csname bibitem#1\endcsname}%
\let\auto@bib@innerbib\@empty
%</preamble>
\bibitem [{\citenamefont {Hu}\ \emph {et~al.}(2000)\citenamefont {Hu},
  \citenamefont {Barkana},\ and\ \citenamefont {Gruzinov}}]{Hu:2000ke}%
  \BibitemOpen
  \bibfield  {author} {\bibinfo {author} {\bibfnamefont {W.}~\bibnamefont
  {Hu}}, \bibinfo {author} {\bibfnamefont {R.}~\bibnamefont {Barkana}}, \ and\
  \bibinfo {author} {\bibfnamefont {A.}~\bibnamefont {Gruzinov}},\ }\href
  {\doibase 10.1103/PhysRevLett.85.1158} {\bibfield  {journal} {\bibinfo
  {journal} {Phys. Rev. Lett.}\ }\textbf {\bibinfo {volume} {85}},\ \bibinfo
  {pages} {1158} (\bibinfo {year} {2000})},\ \Eprint
  {http://arxiv.org/abs/astro-ph/0003365} {arXiv:astro-ph/0003365 [astro-ph]}
  \BibitemShut {NoStop}%
%%CITATION = ASTRO-PH/0003365;%%
\bibitem [{\citenamefont {Hui}\ \emph {et~al.}(2017)\citenamefont {Hui},
  \citenamefont {Ostriker}, \citenamefont {Tremaine},\ and\ \citenamefont
  {Witten}}]{Hui:2016ltb}%
  \BibitemOpen
  \bibfield  {author} {\bibinfo {author} {\bibfnamefont {L.}~\bibnamefont
  {Hui}}, \bibinfo {author} {\bibfnamefont {J.~P.}\ \bibnamefont {Ostriker}},
  \bibinfo {author} {\bibfnamefont {S.}~\bibnamefont {Tremaine}}, \ and\
  \bibinfo {author} {\bibfnamefont {E.}~\bibnamefont {Witten}},\ }\href
  {\doibase 10.1103/PhysRevD.95.043541} {\bibfield  {journal} {\bibinfo
  {journal} {Phys. Rev.}\ }\textbf {\bibinfo {volume} {D95}},\ \bibinfo {pages}
  {043541} (\bibinfo {year} {2017})},\ \Eprint
  {http://arxiv.org/abs/1610.08297} {arXiv:1610.08297 [astro-ph.CO]}
  \BibitemShut {NoStop}%
%%CITATION = ARXIV:1610.08297;%%
\bibitem [{\citenamefont {Inomata}\ \emph {et~al.}(2017)\citenamefont
  {Inomata}, \citenamefont {Kawasaki}, \citenamefont {Mukaida}, \citenamefont
  {Tada},\ and\ \citenamefont {Yanagida}}]{Inomata:2017okj}%
  \BibitemOpen
  \bibfield  {author} {\bibinfo {author} {\bibfnamefont {K.}~\bibnamefont
  {Inomata}}, \bibinfo {author} {\bibfnamefont {M.}~\bibnamefont {Kawasaki}},
  \bibinfo {author} {\bibfnamefont {K.}~\bibnamefont {Mukaida}}, \bibinfo
  {author} {\bibfnamefont {Y.}~\bibnamefont {Tada}}, \ and\ \bibinfo {author}
  {\bibfnamefont {T.~T.}\ \bibnamefont {Yanagida}},\ }\href {\doibase
  10.1103/PhysRevD.96.043504} {\bibfield  {journal} {\bibinfo  {journal} {Phys.
  Rev.}\ }\textbf {\bibinfo {volume} {D96}},\ \bibinfo {pages} {043504}
  (\bibinfo {year} {2017})},\ \Eprint {http://arxiv.org/abs/1701.02544}
  {arXiv:1701.02544 [astro-ph.CO]} \BibitemShut {NoStop}%
%%CITATION = ARXIV:1701.02544;%%
\bibitem [{\citenamefont {Verde}\ \emph {et~al.}(2019)\citenamefont {Verde},
  \citenamefont {Treu},\ and\ \citenamefont {Riess}}]{Verde:2019ivm}%
  \BibitemOpen
  \bibfield  {author} {\bibinfo {author} {\bibfnamefont {L.}~\bibnamefont
  {Verde}}, \bibinfo {author} {\bibfnamefont {T.}~\bibnamefont {Treu}}, \ and\
  \bibinfo {author} {\bibfnamefont {A.~G.}\ \bibnamefont {Riess}},\ }in\ \href
  {\doibase 10.1038/s41550-019-0902-0} {\emph {\bibinfo {booktitle} {{Nature
  Astronomy 2019}}}}\ (\bibinfo {year} {2019})\ \Eprint
  {http://arxiv.org/abs/1907.10625} {arXiv:1907.10625 [astro-ph.CO]}
  \BibitemShut {NoStop}%
%%CITATION = ARXIV:1907.10625;%%
\bibitem [{\citenamefont {Riess}\ \emph {et~al.}(2016)\citenamefont {Riess}
  \emph {et~al.}}]{Riess:2016jrr}%
  \BibitemOpen
  \bibfield  {author} {\bibinfo {author} {\bibfnamefont {A.~G.}\ \bibnamefont
  {Riess}} \emph {et~al.},\ }\href {\doibase 10.3847/0004-637X/826/1/56}
  {\bibfield  {journal} {\bibinfo  {journal} {Astrophys. J.}\ }\textbf
  {\bibinfo {volume} {826}},\ \bibinfo {pages} {56} (\bibinfo {year} {2016})},\
  \Eprint {http://arxiv.org/abs/1604.01424} {arXiv:1604.01424 [astro-ph.CO]}
  \BibitemShut {NoStop}%
%%CITATION = ARXIV:1604.01424;%%
\bibitem [{\citenamefont {Riess}\ \emph {et~al.}(2018)\citenamefont {Riess}
  \emph {et~al.}}]{Riess:2018byc}%
  \BibitemOpen
  \bibfield  {author} {\bibinfo {author} {\bibfnamefont {A.~G.}\ \bibnamefont
  {Riess}} \emph {et~al.},\ }\href {\doibase 10.3847/1538-4357/aac82e}
  {\bibfield  {journal} {\bibinfo  {journal} {Astrophys. J.}\ }\textbf
  {\bibinfo {volume} {861}},\ \bibinfo {pages} {126} (\bibinfo {year}
  {2018})},\ \Eprint {http://arxiv.org/abs/1804.10655} {arXiv:1804.10655
  [astro-ph.CO]} \BibitemShut {NoStop}%
%%CITATION = ARXIV:1804.10655;%%
\bibitem [{\citenamefont {Bonvin}\ \emph {et~al.}(2017)\citenamefont {Bonvin}
  \emph {et~al.}}]{Bonvin:2016crt}%
  \BibitemOpen
  \bibfield  {author} {\bibinfo {author} {\bibfnamefont {V.}~\bibnamefont
  {Bonvin}} \emph {et~al.},\ }\href {\doibase 10.1093/mnras/stw3006} {\bibfield
   {journal} {\bibinfo  {journal} {Mon. Not. Roy. Astron. Soc.}\ }\textbf
  {\bibinfo {volume} {465}},\ \bibinfo {pages} {4914} (\bibinfo {year}
  {2017})},\ \Eprint {http://arxiv.org/abs/1607.01790} {arXiv:1607.01790
  [astro-ph.CO]} \BibitemShut {NoStop}%
%%CITATION = ARXIV:1607.01790;%%
\bibitem [{\citenamefont {Birrer}\ \emph {et~al.}(2019)\citenamefont {Birrer}
  \emph {et~al.}}]{Birrer:2018vtm}%
  \BibitemOpen
  \bibfield  {author} {\bibinfo {author} {\bibfnamefont {S.}~\bibnamefont
  {Birrer}} \emph {et~al.},\ }\href {\doibase 10.1093/mnras/stz200} {\bibfield
  {journal} {\bibinfo  {journal} {Mon. Not. Roy. Astron. Soc.}\ }\textbf
  {\bibinfo {volume} {484}},\ \bibinfo {pages} {4726} (\bibinfo {year}
  {2019})},\ \Eprint {http://arxiv.org/abs/1809.01274} {arXiv:1809.01274
  [astro-ph.CO]} \BibitemShut {NoStop}%
%%CITATION = ARXIV:1809.01274;%%
\bibitem [{\citenamefont {Aghanim}\ \emph {et~al.}(2018)\citenamefont {Aghanim}
  \emph {et~al.}}]{Aghanim:2018eyx}%
  \BibitemOpen
  \bibfield  {author} {\bibinfo {author} {\bibfnamefont {N.}~\bibnamefont
  {Aghanim}} \emph {et~al.} (\bibinfo {collaboration} {Planck}),\ }\href@noop
  {} {\  (\bibinfo {year} {2018})},\ \Eprint {http://arxiv.org/abs/1807.06209}
  {arXiv:1807.06209 [astro-ph.CO]} \BibitemShut {NoStop}%
%%CITATION = ARXIV:1807.06209;%%
\bibitem [{\citenamefont {Efstathiou}(2014)}]{Efstathiou:2013via}%
  \BibitemOpen
  \bibfield  {author} {\bibinfo {author} {\bibfnamefont {G.}~\bibnamefont
  {Efstathiou}},\ }\href {\doibase 10.1093/mnras/stu278} {\bibfield  {journal}
  {\bibinfo  {journal} {Mon. Not. Roy. Astron. Soc.}\ }\textbf {\bibinfo
  {volume} {440}},\ \bibinfo {pages} {1138} (\bibinfo {year} {2014})},\ \Eprint
  {http://arxiv.org/abs/1311.3461} {arXiv:1311.3461 [astro-ph.CO]} \BibitemShut
  {NoStop}%
%%CITATION = ARXIV:1311.3461;%%
\bibitem [{\citenamefont {Freedman}(2017)}]{Freedman:2017yms}%
  \BibitemOpen
  \bibfield  {author} {\bibinfo {author} {\bibfnamefont {W.~L.}\ \bibnamefont
  {Freedman}},\ }\href {\doibase 10.1038/s41550-017-0121} {\bibfield  {journal}
  {\bibinfo  {journal} {Nat. Astron.}\ }\textbf {\bibinfo {volume} {1}},\
  \bibinfo {pages} {0121} (\bibinfo {year} {2017})},\ \Eprint
  {http://arxiv.org/abs/1706.02739} {arXiv:1706.02739 [astro-ph.CO]}
  \BibitemShut {NoStop}%
%%CITATION = ARXIV:1706.02739;%%
\bibitem [{\citenamefont {Rameez}\ and\ \citenamefont
  {Sarkar}(2019)}]{Rameez:2019wdt}%
  \BibitemOpen
  \bibfield  {author} {\bibinfo {author} {\bibfnamefont {M.}~\bibnamefont
  {Rameez}}\ and\ \bibinfo {author} {\bibfnamefont {S.}~\bibnamefont
  {Sarkar}},\ }\href@noop {} {\  (\bibinfo {year} {2019})},\ \Eprint
  {http://arxiv.org/abs/1911.06456} {arXiv:1911.06456 [astro-ph.CO]}
  \BibitemShut {NoStop}%
%%CITATION = ARXIV:1911.06456;%%
\bibitem [{\citenamefont {Vattis}\ \emph {et~al.}(2019)\citenamefont {Vattis},
  \citenamefont {Koushiappas},\ and\ \citenamefont {Loeb}}]{Vattis:2019efj}%
  \BibitemOpen
  \bibfield  {author} {\bibinfo {author} {\bibfnamefont {K.}~\bibnamefont
  {Vattis}}, \bibinfo {author} {\bibfnamefont {S.~M.}\ \bibnamefont
  {Koushiappas}}, \ and\ \bibinfo {author} {\bibfnamefont {A.}~\bibnamefont
  {Loeb}},\ }\href {\doibase 10.1103/PhysRevD.99.121302} {\bibfield  {journal}
  {\bibinfo  {journal} {Phys. Rev.}\ }\textbf {\bibinfo {volume} {D99}},\
  \bibinfo {pages} {121302} (\bibinfo {year} {2019})},\ \Eprint
  {http://arxiv.org/abs/1903.06220} {arXiv:1903.06220 [astro-ph.CO]}
  \BibitemShut {NoStop}%
%%CITATION = ARXIV:1903.06220;%%
\bibitem [{\citenamefont {Choi}\ \emph
  {et~al.}(2020{\natexlab{a}})\citenamefont {Choi}, \citenamefont {Suzuki},\
  and\ \citenamefont {Yanagida}}]{Choi:2019jck}%
  \BibitemOpen
  \bibfield  {author} {\bibinfo {author} {\bibfnamefont {G.}~\bibnamefont
  {Choi}}, \bibinfo {author} {\bibfnamefont {M.}~\bibnamefont {Suzuki}}, \ and\
  \bibinfo {author} {\bibfnamefont {T.~T.}\ \bibnamefont {Yanagida}},\ }\href
  {\doibase 10.1016/j.physletb.2020.135408} {\bibfield  {journal} {\bibinfo
  {journal} {Phys. Lett. B}\ }\textbf {\bibinfo {volume} {805}},\ \bibinfo
  {pages} {135408} (\bibinfo {year} {2020}{\natexlab{a}})},\ \Eprint
  {http://arxiv.org/abs/1910.00459} {arXiv:1910.00459 [hep-ph]} \BibitemShut
  {NoStop}%
\bibitem [{\citenamefont {Feng}\ \emph {et~al.}(2003)\citenamefont {Feng},
  \citenamefont {Rajaraman},\ and\ \citenamefont {Takayama}}]{Feng:2003xh}%
  \BibitemOpen
  \bibfield  {author} {\bibinfo {author} {\bibfnamefont {J.~L.}\ \bibnamefont
  {Feng}}, \bibinfo {author} {\bibfnamefont {A.}~\bibnamefont {Rajaraman}}, \
  and\ \bibinfo {author} {\bibfnamefont {F.}~\bibnamefont {Takayama}},\ }\href
  {\doibase 10.1103/PhysRevLett.91.011302} {\bibfield  {journal} {\bibinfo
  {journal} {Phys. Rev. Lett.}\ }\textbf {\bibinfo {volume} {91}},\ \bibinfo
  {pages} {011302} (\bibinfo {year} {2003})},\ \Eprint
  {http://arxiv.org/abs/hep-ph/0302215} {arXiv:hep-ph/0302215 [hep-ph]}
  \BibitemShut {NoStop}%
%%CITATION = HEP-PH/0302215;%%
\bibitem [{\citenamefont {Bernal}\ \emph {et~al.}(2016)\citenamefont {Bernal},
  \citenamefont {Verde},\ and\ \citenamefont {Riess}}]{Bernal:2016gxb}%
  \BibitemOpen
  \bibfield  {author} {\bibinfo {author} {\bibfnamefont {J.~L.}\ \bibnamefont
  {Bernal}}, \bibinfo {author} {\bibfnamefont {L.}~\bibnamefont {Verde}}, \
  and\ \bibinfo {author} {\bibfnamefont {A.~G.}\ \bibnamefont {Riess}},\ }\href
  {\doibase 10.1088/1475-7516/2016/10/019} {\bibfield  {journal} {\bibinfo
  {journal} {JCAP}\ }\textbf {\bibinfo {volume} {1610}},\ \bibinfo {pages}
  {019} (\bibinfo {year} {2016})},\ \Eprint {http://arxiv.org/abs/1607.05617}
  {arXiv:1607.05617 [astro-ph.CO]} \BibitemShut {NoStop}%
%%CITATION = ARXIV:1607.05617;%%
\bibitem [{\citenamefont {Vagnozzi}(2019)}]{Vagnozzi:2019ezj}%
  \BibitemOpen
  \bibfield  {author} {\bibinfo {author} {\bibfnamefont {S.}~\bibnamefont
  {Vagnozzi}},\ }\href@noop {} {\  (\bibinfo {year} {2019})},\ \Eprint
  {http://arxiv.org/abs/1907.07569} {arXiv:1907.07569 [astro-ph.CO]}
  \BibitemShut {NoStop}%
%%CITATION = ARXIV:1907.07569;%%
\bibitem [{\citenamefont {Karwal}\ and\ \citenamefont
  {Kamionkowski}(2016)}]{Karwal:2016vyq}%
  \BibitemOpen
  \bibfield  {author} {\bibinfo {author} {\bibfnamefont {T.}~\bibnamefont
  {Karwal}}\ and\ \bibinfo {author} {\bibfnamefont {M.}~\bibnamefont
  {Kamionkowski}},\ }\href {\doibase 10.1103/PhysRevD.94.103523} {\bibfield
  {journal} {\bibinfo  {journal} {Phys. Rev.}\ }\textbf {\bibinfo {volume}
  {D94}},\ \bibinfo {pages} {103523} (\bibinfo {year} {2016})},\ \Eprint
  {http://arxiv.org/abs/1608.01309} {arXiv:1608.01309 [astro-ph.CO]}
  \BibitemShut {NoStop}%
%%CITATION = ARXIV:1608.01309;%%
\bibitem [{\citenamefont {Poulin}\ \emph {et~al.}(2019)\citenamefont {Poulin},
  \citenamefont {Smith}, \citenamefont {Karwal},\ and\ \citenamefont
  {Kamionkowski}}]{Poulin:2018cxd}%
  \BibitemOpen
  \bibfield  {author} {\bibinfo {author} {\bibfnamefont {V.}~\bibnamefont
  {Poulin}}, \bibinfo {author} {\bibfnamefont {T.~L.}\ \bibnamefont {Smith}},
  \bibinfo {author} {\bibfnamefont {T.}~\bibnamefont {Karwal}}, \ and\ \bibinfo
  {author} {\bibfnamefont {M.}~\bibnamefont {Kamionkowski}},\ }\href {\doibase
  10.1103/PhysRevLett.122.221301} {\bibfield  {journal} {\bibinfo  {journal}
  {Phys. Rev. Lett.}\ }\textbf {\bibinfo {volume} {122}},\ \bibinfo {pages}
  {221301} (\bibinfo {year} {2019})},\ \Eprint
  {http://arxiv.org/abs/1811.04083} {arXiv:1811.04083 [astro-ph.CO]}
  \BibitemShut {NoStop}%
%%CITATION = ARXIV:1811.04083;%%
\bibitem [{\citenamefont {Alexander}\ and\ \citenamefont
  {McDonough}(2019)}]{Alexander:2019rsc}%
  \BibitemOpen
  \bibfield  {author} {\bibinfo {author} {\bibfnamefont {S.}~\bibnamefont
  {Alexander}}\ and\ \bibinfo {author} {\bibfnamefont {E.}~\bibnamefont
  {McDonough}},\ }\href {\doibase 10.1016/j.physletb.2019.134830} {\bibfield
  {journal} {\bibinfo  {journal} {Phys. Lett.}\ }\textbf {\bibinfo {volume}
  {B797}},\ \bibinfo {pages} {134830} (\bibinfo {year} {2019})},\ \Eprint
  {http://arxiv.org/abs/1904.08912} {arXiv:1904.08912 [astro-ph.CO]}
  \BibitemShut {NoStop}%
%%CITATION = ARXIV:1904.08912;%%
\bibitem [{\citenamefont {Sakstein}\ and\ \citenamefont
  {Trodden}(2019)}]{Sakstein:2019fmf}%
  \BibitemOpen
  \bibfield  {author} {\bibinfo {author} {\bibfnamefont {J.}~\bibnamefont
  {Sakstein}}\ and\ \bibinfo {author} {\bibfnamefont {M.}~\bibnamefont
  {Trodden}},\ }\href@noop {} {\  (\bibinfo {year} {2019})},\ \Eprint
  {http://arxiv.org/abs/1911.11760} {arXiv:1911.11760 [astro-ph.CO]}
  \BibitemShut {NoStop}%
%%CITATION = ARXIV:1911.11760;%%
\bibitem [{\citenamefont {Yang}\ \emph {et~al.}(2018)\citenamefont {Yang},
  \citenamefont {Pan}, \citenamefont {Di~Valentino}, \citenamefont {Nunes},
  \citenamefont {Vagnozzi},\ and\ \citenamefont {Mota}}]{Yang:2018euj}%
  \BibitemOpen
  \bibfield  {author} {\bibinfo {author} {\bibfnamefont {W.}~\bibnamefont
  {Yang}}, \bibinfo {author} {\bibfnamefont {S.}~\bibnamefont {Pan}}, \bibinfo
  {author} {\bibfnamefont {E.}~\bibnamefont {Di~Valentino}}, \bibinfo {author}
  {\bibfnamefont {R.~C.}\ \bibnamefont {Nunes}}, \bibinfo {author}
  {\bibfnamefont {S.}~\bibnamefont {Vagnozzi}}, \ and\ \bibinfo {author}
  {\bibfnamefont {D.~F.}\ \bibnamefont {Mota}},\ }\href {\doibase
  10.1088/1475-7516/2018/09/019} {\bibfield  {journal} {\bibinfo  {journal}
  {JCAP}\ }\textbf {\bibinfo {volume} {1809}},\ \bibinfo {pages} {019}
  (\bibinfo {year} {2018})},\ \Eprint {http://arxiv.org/abs/1805.08252}
  {arXiv:1805.08252 [astro-ph.CO]} \BibitemShut {NoStop}%
%%CITATION = ARXIV:1805.08252;%%
\bibitem [{\citenamefont {Di~Valentino}\ \emph {et~al.}(2019)\citenamefont
  {Di~Valentino}, \citenamefont {Melchiorri}, \citenamefont {Mena},\ and\
  \citenamefont {Vagnozzi}}]{DiValentino:2019ffd}%
  \BibitemOpen
  \bibfield  {author} {\bibinfo {author} {\bibfnamefont {E.}~\bibnamefont
  {Di~Valentino}}, \bibinfo {author} {\bibfnamefont {A.}~\bibnamefont
  {Melchiorri}}, \bibinfo {author} {\bibfnamefont {O.}~\bibnamefont {Mena}}, \
  and\ \bibinfo {author} {\bibfnamefont {S.}~\bibnamefont {Vagnozzi}},\
  }\href@noop {} {\  (\bibinfo {year} {2019})},\ \Eprint
  {http://arxiv.org/abs/1908.04281} {arXiv:1908.04281 [astro-ph.CO]}
  \BibitemShut {NoStop}%
%%CITATION = ARXIV:1908.04281;%%
\bibitem [{\citenamefont {Moore}\ \emph
  {et~al.}(1999{\natexlab{a}})\citenamefont {Moore}, \citenamefont {Quinn},
  \citenamefont {Governato}, \citenamefont {Stadel},\ and\ \citenamefont
  {Lake}}]{Moore:1999gc}%
  \BibitemOpen
  \bibfield  {author} {\bibinfo {author} {\bibfnamefont {B.}~\bibnamefont
  {Moore}}, \bibinfo {author} {\bibfnamefont {T.~R.}\ \bibnamefont {Quinn}},
  \bibinfo {author} {\bibfnamefont {F.}~\bibnamefont {Governato}}, \bibinfo
  {author} {\bibfnamefont {J.}~\bibnamefont {Stadel}}, \ and\ \bibinfo {author}
  {\bibfnamefont {G.}~\bibnamefont {Lake}},\ }\href {\doibase
  10.1046/j.1365-8711.1999.03039.x} {\bibfield  {journal} {\bibinfo  {journal}
  {Mon. Not. Roy. Astron. Soc.}\ }\textbf {\bibinfo {volume} {310}},\ \bibinfo
  {pages} {1147} (\bibinfo {year} {1999}{\natexlab{a}})},\ \Eprint
  {http://arxiv.org/abs/astro-ph/9903164} {arXiv:astro-ph/9903164 [astro-ph]}
  \BibitemShut {NoStop}%
%%CITATION = ASTRO-PH/9903164;%%
\bibitem [{\citenamefont {Moore}\ \emph
  {et~al.}(1999{\natexlab{b}})\citenamefont {Moore}, \citenamefont {Ghigna},
  \citenamefont {Governato}, \citenamefont {Lake}, \citenamefont {Quinn},
  \citenamefont {Stadel},\ and\ \citenamefont {Tozzi}}]{Moore:1999nt}%
  \BibitemOpen
  \bibfield  {author} {\bibinfo {author} {\bibfnamefont {B.}~\bibnamefont
  {Moore}}, \bibinfo {author} {\bibfnamefont {S.}~\bibnamefont {Ghigna}},
  \bibinfo {author} {\bibfnamefont {F.}~\bibnamefont {Governato}}, \bibinfo
  {author} {\bibfnamefont {G.}~\bibnamefont {Lake}}, \bibinfo {author}
  {\bibfnamefont {T.~R.}\ \bibnamefont {Quinn}}, \bibinfo {author}
  {\bibfnamefont {J.}~\bibnamefont {Stadel}}, \ and\ \bibinfo {author}
  {\bibfnamefont {P.}~\bibnamefont {Tozzi}},\ }\href {\doibase 10.1086/312287}
  {\bibfield  {journal} {\bibinfo  {journal} {Astrophys. J.}\ }\textbf
  {\bibinfo {volume} {524}},\ \bibinfo {pages} {L19} (\bibinfo {year}
  {1999}{\natexlab{b}})},\ \Eprint {http://arxiv.org/abs/astro-ph/9907411}
  {arXiv:astro-ph/9907411 [astro-ph]} \BibitemShut {NoStop}%
%%CITATION = ASTRO-PH/9907411;%%
\bibitem [{\citenamefont {Kim}\ \emph {et~al.}(2018)\citenamefont {Kim},
  \citenamefont {Peter},\ and\ \citenamefont {Hargis}}]{Kim:2017iwr}%
  \BibitemOpen
  \bibfield  {author} {\bibinfo {author} {\bibfnamefont {S.~Y.}\ \bibnamefont
  {Kim}}, \bibinfo {author} {\bibfnamefont {A.~H.~G.}\ \bibnamefont {Peter}}, \
  and\ \bibinfo {author} {\bibfnamefont {J.~R.}\ \bibnamefont {Hargis}},\
  }\href {\doibase 10.1103/PhysRevLett.121.211302} {\bibfield  {journal}
  {\bibinfo  {journal} {Phys. Rev. Lett.}\ }\textbf {\bibinfo {volume} {121}},\
  \bibinfo {pages} {211302} (\bibinfo {year} {2018})},\ \Eprint
  {http://arxiv.org/abs/1711.06267} {arXiv:1711.06267 [astro-ph.CO]}
  \BibitemShut {NoStop}%
%%CITATION = ARXIV:1711.06267;%%
\bibitem [{\citenamefont {Boylan-Kolchin}\ \emph {et~al.}(2011)\citenamefont
  {Boylan-Kolchin}, \citenamefont {Bullock},\ and\ \citenamefont
  {Kaplinghat}}]{Boylan_Kolchin_2011}%
  \BibitemOpen
  \bibfield  {author} {\bibinfo {author} {\bibfnamefont {M.}~\bibnamefont
  {Boylan-Kolchin}}, \bibinfo {author} {\bibfnamefont {J.~S.}\ \bibnamefont
  {Bullock}}, \ and\ \bibinfo {author} {\bibfnamefont {M.}~\bibnamefont
  {Kaplinghat}},\ }\href {\doibase 10.1111/j.1745-3933.2011.01074.x} {\bibfield
   {journal} {\bibinfo  {journal} {Monthly Notices of the Royal Astronomical
  Society: Letters}\ }\textbf {\bibinfo {volume} {415}},\ \bibinfo {pages}
  {L40} (\bibinfo {year} {2011})}\BibitemShut {NoStop}%
\bibitem [{\citenamefont {Weinberg}\ \emph {et~al.}(2015)\citenamefont
  {Weinberg}, \citenamefont {Bullock}, \citenamefont {Governato}, \citenamefont
  {Kuzio~de Naray},\ and\ \citenamefont {Peter}}]{Weinberg:2013aya}%
  \BibitemOpen
  \bibfield  {author} {\bibinfo {author} {\bibfnamefont {D.~H.}\ \bibnamefont
  {Weinberg}}, \bibinfo {author} {\bibfnamefont {J.~S.}\ \bibnamefont
  {Bullock}}, \bibinfo {author} {\bibfnamefont {F.}~\bibnamefont {Governato}},
  \bibinfo {author} {\bibfnamefont {R.}~\bibnamefont {Kuzio~de Naray}}, \ and\
  \bibinfo {author} {\bibfnamefont {A.~H.~G.}\ \bibnamefont {Peter}},\
  }\bibfield  {booktitle} {\emph {\bibinfo {booktitle} {{Sackler Colloquium:
  Dark Matter Universe: On the Threshhold of Discovery Irvine, USA, October
  18-20, 2012}}},\ }\href {\doibase 10.1073/pnas.1308716112} {\bibfield
  {journal} {\bibinfo  {journal} {Proc. Nat. Acad. Sci.}\ }\textbf {\bibinfo
  {volume} {112}},\ \bibinfo {pages} {12249} (\bibinfo {year} {2015})},\
  \Eprint {http://arxiv.org/abs/1306.0913} {arXiv:1306.0913 [astro-ph.CO]}
  \BibitemShut {NoStop}%
%%CITATION = ARXIV:1306.0913;%%
\bibitem [{\citenamefont {Navarro}\ \emph {et~al.}(1997)\citenamefont
  {Navarro}, \citenamefont {Frenk},\ and\ \citenamefont
  {White}}]{Navarro:1996gj}%
  \BibitemOpen
  \bibfield  {author} {\bibinfo {author} {\bibfnamefont {J.~F.}\ \bibnamefont
  {Navarro}}, \bibinfo {author} {\bibfnamefont {C.~S.}\ \bibnamefont {Frenk}},
  \ and\ \bibinfo {author} {\bibfnamefont {S.~D.~M.}\ \bibnamefont {White}},\
  }\href {\doibase 10.1086/304888} {\bibfield  {journal} {\bibinfo  {journal}
  {Astrophys. J.}\ }\textbf {\bibinfo {volume} {490}},\ \bibinfo {pages} {493}
  (\bibinfo {year} {1997})},\ \Eprint {http://arxiv.org/abs/astro-ph/9611107}
  {arXiv:astro-ph/9611107 [astro-ph]} \BibitemShut {NoStop}%
%%CITATION = ASTRO-PH/9611107;%%
\bibitem [{\citenamefont {Fukushige}\ and\ \citenamefont
  {Makino}(1997)}]{Fukushige:1996nr}%
  \BibitemOpen
  \bibfield  {author} {\bibinfo {author} {\bibfnamefont {T.}~\bibnamefont
  {Fukushige}}\ and\ \bibinfo {author} {\bibfnamefont {J.}~\bibnamefont
  {Makino}},\ }\href {\doibase 10.1086/310516} {\bibfield  {journal} {\bibinfo
  {journal} {Astrophys. J.}\ }\textbf {\bibinfo {volume} {477}},\ \bibinfo
  {pages} {L9} (\bibinfo {year} {1997})},\ \Eprint
  {http://arxiv.org/abs/astro-ph/9610005} {arXiv:astro-ph/9610005 [astro-ph]}
  \BibitemShut {NoStop}%
%%CITATION = ASTRO-PH/9610005;%%
\bibitem [{\citenamefont {Ishiyama}\ \emph {et~al.}(2013)\citenamefont
  {Ishiyama}, \citenamefont {Makino}, \citenamefont {Portegies~Zwart},
  \citenamefont {Groen}, \citenamefont {Nitadori}, \citenamefont {Rieder},
  \citenamefont {de~Laat}, \citenamefont {McMillan}, \citenamefont {Hiraki},\
  and\ \citenamefont {Harfst}}]{Ishiyama:2011af}%
  \BibitemOpen
  \bibfield  {author} {\bibinfo {author} {\bibfnamefont {T.}~\bibnamefont
  {Ishiyama}}, \bibinfo {author} {\bibfnamefont {J.}~\bibnamefont {Makino}},
  \bibinfo {author} {\bibfnamefont {S.}~\bibnamefont {Portegies~Zwart}},
  \bibinfo {author} {\bibfnamefont {D.}~\bibnamefont {Groen}}, \bibinfo
  {author} {\bibfnamefont {K.}~\bibnamefont {Nitadori}}, \bibinfo {author}
  {\bibfnamefont {S.}~\bibnamefont {Rieder}}, \bibinfo {author} {\bibfnamefont
  {C.}~\bibnamefont {de~Laat}}, \bibinfo {author} {\bibfnamefont
  {S.}~\bibnamefont {McMillan}}, \bibinfo {author} {\bibfnamefont
  {K.}~\bibnamefont {Hiraki}}, \ and\ \bibinfo {author} {\bibfnamefont
  {S.}~\bibnamefont {Harfst}},\ }\href {\doibase 10.1088/0004-637X/767/2/146}
  {\bibfield  {journal} {\bibinfo  {journal} {Astrophys. J.}\ }\textbf
  {\bibinfo {volume} {767}},\ \bibinfo {pages} {146} (\bibinfo {year}
  {2013})},\ \Eprint {http://arxiv.org/abs/1101.2020} {arXiv:1101.2020
  [astro-ph.CO]} \BibitemShut {NoStop}%
%%CITATION = ARXIV:1101.2020;%%
\bibitem [{\citenamefont {Borriello}\ and\ \citenamefont
  {Salucci}(2001)}]{Borriello:2000rv}%
  \BibitemOpen
  \bibfield  {author} {\bibinfo {author} {\bibfnamefont {A.}~\bibnamefont
  {Borriello}}\ and\ \bibinfo {author} {\bibfnamefont {P.}~\bibnamefont
  {Salucci}},\ }\href {\doibase 10.1046/j.1365-8711.2001.04077.x} {\bibfield
  {journal} {\bibinfo  {journal} {Mon. Not. Roy. Astron. Soc.}\ }\textbf
  {\bibinfo {volume} {323}},\ \bibinfo {pages} {285} (\bibinfo {year}
  {2001})},\ \Eprint {http://arxiv.org/abs/astro-ph/0001082}
  {arXiv:astro-ph/0001082 [astro-ph]} \BibitemShut {NoStop}%
%%CITATION = ASTRO-PH/0001082;%%
\bibitem [{\citenamefont {Gilmore}\ \emph {et~al.}(2007)\citenamefont
  {Gilmore}, \citenamefont {Wilkinson}, \citenamefont {Wyse}, \citenamefont
  {Kleyna}, \citenamefont {Koch}, \citenamefont {Evans},\ and\ \citenamefont
  {Grebel}}]{Gilmore:2007fy}%
  \BibitemOpen
  \bibfield  {author} {\bibinfo {author} {\bibfnamefont {G.}~\bibnamefont
  {Gilmore}}, \bibinfo {author} {\bibfnamefont {M.~I.}\ \bibnamefont
  {Wilkinson}}, \bibinfo {author} {\bibfnamefont {R.~F.~G.}\ \bibnamefont
  {Wyse}}, \bibinfo {author} {\bibfnamefont {J.~T.}\ \bibnamefont {Kleyna}},
  \bibinfo {author} {\bibfnamefont {A.}~\bibnamefont {Koch}}, \bibinfo {author}
  {\bibfnamefont {N.~W.}\ \bibnamefont {Evans}}, \ and\ \bibinfo {author}
  {\bibfnamefont {E.~K.}\ \bibnamefont {Grebel}},\ }\href {\doibase
  10.1086/518025} {\bibfield  {journal} {\bibinfo  {journal} {Astrophys. J.}\
  }\textbf {\bibinfo {volume} {663}},\ \bibinfo {pages} {948} (\bibinfo {year}
  {2007})},\ \Eprint {http://arxiv.org/abs/astro-ph/0703308}
  {arXiv:astro-ph/0703308 [ASTRO-PH]} \BibitemShut {NoStop}%
%%CITATION = ASTRO-PH/0703308;%%
\bibitem [{\citenamefont {Oh}\ \emph {et~al.}(2008)\citenamefont {Oh},
  \citenamefont {de~Blok}, \citenamefont {Walter}, \citenamefont {Brinks},\
  and\ \citenamefont {Kennicutt}}]{Oh:2008ww}%
  \BibitemOpen
  \bibfield  {author} {\bibinfo {author} {\bibfnamefont {S.-H.}\ \bibnamefont
  {Oh}}, \bibinfo {author} {\bibfnamefont {W.~J.~G.}\ \bibnamefont {de~Blok}},
  \bibinfo {author} {\bibfnamefont {F.}~\bibnamefont {Walter}}, \bibinfo
  {author} {\bibfnamefont {E.}~\bibnamefont {Brinks}}, \ and\ \bibinfo {author}
  {\bibfnamefont {R.~C.}\ \bibnamefont {Kennicutt}, \bibfnamefont {Jr}},\
  }\href {\doibase 10.1088/0004-6256/136/6/2761} {\bibfield  {journal}
  {\bibinfo  {journal} {Astron. J.}\ }\textbf {\bibinfo {volume} {136}},\
  \bibinfo {pages} {2761} (\bibinfo {year} {2008})},\ \Eprint
  {http://arxiv.org/abs/0810.2119} {arXiv:0810.2119 [astro-ph]} \BibitemShut
  {NoStop}%
%%CITATION = ARXIV:0810.2119;%%
\bibitem [{\citenamefont {de~Blok}(2010)}]{deBlok:2009sp}%
  \BibitemOpen
  \bibfield  {author} {\bibinfo {author} {\bibfnamefont {W.~J.~G.}\
  \bibnamefont {de~Blok}},\ }\href {\doibase 10.1155/2010/789293} {\bibfield
  {journal} {\bibinfo  {journal} {Adv. Astron.}\ }\textbf {\bibinfo {volume}
  {2010}},\ \bibinfo {pages} {789293} (\bibinfo {year} {2010})},\ \Eprint
  {http://arxiv.org/abs/0910.3538} {arXiv:0910.3538 [astro-ph.CO]} \BibitemShut
  {NoStop}%
%%CITATION = ARXIV:0910.3538;%%
\bibitem [{\citenamefont {Destri}\ \emph {et~al.}(2013)\citenamefont {Destri},
  \citenamefont {de~Vega},\ and\ \citenamefont {Sanchez}}]{Destri:2012yn}%
  \BibitemOpen
  \bibfield  {author} {\bibinfo {author} {\bibfnamefont {C.}~\bibnamefont
  {Destri}}, \bibinfo {author} {\bibfnamefont {H.~J.}\ \bibnamefont {de~Vega}},
  \ and\ \bibinfo {author} {\bibfnamefont {N.~G.}\ \bibnamefont {Sanchez}},\
  }\href {\doibase 10.1016/j.newast.2012.12.003} {\bibfield  {journal}
  {\bibinfo  {journal} {New Astron.}\ }\textbf {\bibinfo {volume} {22}},\
  \bibinfo {pages} {39} (\bibinfo {year} {2013})},\ \Eprint
  {http://arxiv.org/abs/1204.3090} {arXiv:1204.3090 [astro-ph.CO]} \BibitemShut
  {NoStop}%
%%CITATION = ARXIV:1204.3090;%%
\bibitem [{\citenamefont {Domcke}\ and\ \citenamefont
  {Urbano}(2015)}]{Domcke:2014kla}%
  \BibitemOpen
  \bibfield  {author} {\bibinfo {author} {\bibfnamefont {V.}~\bibnamefont
  {Domcke}}\ and\ \bibinfo {author} {\bibfnamefont {A.}~\bibnamefont
  {Urbano}},\ }\href {\doibase 10.1088/1475-7516/2015/01/002} {\bibfield
  {journal} {\bibinfo  {journal} {JCAP}\ }\textbf {\bibinfo {volume} {1501}},\
  \bibinfo {pages} {002} (\bibinfo {year} {2015})},\ \Eprint
  {http://arxiv.org/abs/1409.3167} {arXiv:1409.3167 [hep-ph]} \BibitemShut
  {NoStop}%
%%CITATION = ARXIV:1409.3167;%%
\bibitem [{\citenamefont {Alexander}\ and\ \citenamefont
  {Cormack}(2017)}]{Alexander:2016glq}%
  \BibitemOpen
  \bibfield  {author} {\bibinfo {author} {\bibfnamefont {S.}~\bibnamefont
  {Alexander}}\ and\ \bibinfo {author} {\bibfnamefont {S.}~\bibnamefont
  {Cormack}},\ }\href {\doibase 10.1088/1475-7516/2017/04/005} {\bibfield
  {journal} {\bibinfo  {journal} {JCAP}\ }\textbf {\bibinfo {volume} {1704}},\
  \bibinfo {pages} {005} (\bibinfo {year} {2017})},\ \Eprint
  {http://arxiv.org/abs/1607.08621} {arXiv:1607.08621 [astro-ph.CO]}
  \BibitemShut {NoStop}%
%%CITATION = ARXIV:1607.08621;%%
\bibitem [{\citenamefont {Randall}\ \emph {et~al.}(2017)\citenamefont
  {Randall}, \citenamefont {Scholtz},\ and\ \citenamefont
  {Unwin}}]{Randall:2016bqw}%
  \BibitemOpen
  \bibfield  {author} {\bibinfo {author} {\bibfnamefont {L.}~\bibnamefont
  {Randall}}, \bibinfo {author} {\bibfnamefont {J.}~\bibnamefont {Scholtz}}, \
  and\ \bibinfo {author} {\bibfnamefont {J.}~\bibnamefont {Unwin}},\ }\href
  {\doibase 10.1093/mnras/stx161} {\bibfield  {journal} {\bibinfo  {journal}
  {Mon. Not. Roy. Astron. Soc.}\ }\textbf {\bibinfo {volume} {467}},\ \bibinfo
  {pages} {1515} (\bibinfo {year} {2017})},\ \Eprint
  {http://arxiv.org/abs/1611.04590} {arXiv:1611.04590 [astro-ph.GA]}
  \BibitemShut {NoStop}%
%%CITATION = ARXIV:1611.04590;%%
\bibitem [{\citenamefont {Giraud}\ and\ \citenamefont
  {Peschanski}(2019)}]{Giraud:2018gxl}%
  \BibitemOpen
  \bibfield  {author} {\bibinfo {author} {\bibfnamefont {B.~G.}\ \bibnamefont
  {Giraud}}\ and\ \bibinfo {author} {\bibfnamefont {R.}~\bibnamefont
  {Peschanski}},\ }\href {\doibase 10.1088/1402-4896/ab1959} {\bibfield
  {journal} {\bibinfo  {journal} {Phys. Scripta}\ }\textbf {\bibinfo {volume}
  {94}},\ \bibinfo {pages} {085003} (\bibinfo {year} {2019})},\ \Eprint
  {http://arxiv.org/abs/1806.07283} {arXiv:1806.07283 [hep-th]} \BibitemShut
  {NoStop}%
%%CITATION = ARXIV:1806.07283;%%
\bibitem [{\citenamefont {Savchenko}\ and\ \citenamefont
  {Rudakovskyi}(2019)}]{Savchenko:2019qnn}%
  \BibitemOpen
  \bibfield  {author} {\bibinfo {author} {\bibfnamefont {D.}~\bibnamefont
  {Savchenko}}\ and\ \bibinfo {author} {\bibfnamefont {A.}~\bibnamefont
  {Rudakovskyi}},\ }\href {\doibase 10.1093/mnras/stz1573} {\bibfield
  {journal} {\bibinfo  {journal} {Mon. Not. Roy. Astron. Soc.}\ }\textbf
  {\bibinfo {volume} {487}},\ \bibinfo {pages} {5711} (\bibinfo {year}
  {2019})},\ \Eprint {http://arxiv.org/abs/1903.01862} {arXiv:1903.01862
  [astro-ph.CO]} \BibitemShut {NoStop}%
%%CITATION = ARXIV:1903.01862;%%
\bibitem [{\citenamefont {Viel}\ \emph {et~al.}(2013)\citenamefont {Viel},
  \citenamefont {Becker}, \citenamefont {Bolton},\ and\ \citenamefont
  {Haehnelt}}]{Viel:2013apy}%
  \BibitemOpen
  \bibfield  {author} {\bibinfo {author} {\bibfnamefont {M.}~\bibnamefont
  {Viel}}, \bibinfo {author} {\bibfnamefont {G.~D.}\ \bibnamefont {Becker}},
  \bibinfo {author} {\bibfnamefont {J.~S.}\ \bibnamefont {Bolton}}, \ and\
  \bibinfo {author} {\bibfnamefont {M.~G.}\ \bibnamefont {Haehnelt}},\ }\href
  {\doibase 10.1103/PhysRevD.88.043502} {\bibfield  {journal} {\bibinfo
  {journal} {Phys. Rev.}\ }\textbf {\bibinfo {volume} {D88}},\ \bibinfo {pages}
  {043502} (\bibinfo {year} {2013})},\ \Eprint {http://arxiv.org/abs/1306.2314}
  {arXiv:1306.2314 [astro-ph.CO]} \BibitemShut {NoStop}%
%%CITATION = ARXIV:1306.2314;%%
\bibitem [{\citenamefont {Ir{\v s}i{\v c}}\ \emph {et~al.}(2017)\citenamefont
  {Ir{\v s}i{\v c}} \emph {et~al.}}]{Irsic:2017ixq}%
  \BibitemOpen
  \bibfield  {author} {\bibinfo {author} {\bibfnamefont {V.}~\bibnamefont
  {Ir{\v s}i{\v c}}} \emph {et~al.},\ }\href {\doibase
  10.1103/PhysRevD.96.023522} {\bibfield  {journal} {\bibinfo  {journal} {Phys.
  Rev.}\ }\textbf {\bibinfo {volume} {D96}},\ \bibinfo {pages} {023522}
  (\bibinfo {year} {2017})},\ \Eprint {http://arxiv.org/abs/1702.01764}
  {arXiv:1702.01764 [astro-ph.CO]} \BibitemShut {NoStop}%
%%CITATION = ARXIV:1702.01764;%%
\bibitem [{\citenamefont {Nakayama}\ \emph {et~al.}(2011)\citenamefont
  {Nakayama}, \citenamefont {Takahashi},\ and\ \citenamefont
  {Yanagida}}]{Nakayama:2011dj}%
  \BibitemOpen
  \bibfield  {author} {\bibinfo {author} {\bibfnamefont {K.}~\bibnamefont
  {Nakayama}}, \bibinfo {author} {\bibfnamefont {F.}~\bibnamefont {Takahashi}},
  \ and\ \bibinfo {author} {\bibfnamefont {T.~T.}\ \bibnamefont {Yanagida}},\
  }\href {\doibase 10.1016/j.physletb.2011.04.035} {\bibfield  {journal}
  {\bibinfo  {journal} {Phys. Lett.}\ }\textbf {\bibinfo {volume} {B699}},\
  \bibinfo {pages} {360} (\bibinfo {year} {2011})},\ \Eprint
  {http://arxiv.org/abs/1102.4688} {arXiv:1102.4688 [hep-ph]} \BibitemShut
  {NoStop}%
%%CITATION = ARXIV:1102.4688;%%
\bibitem [{\citenamefont {Nakayama}\ \emph {et~al.}(2019)\citenamefont
  {Nakayama}, \citenamefont {Takahashi},\ and\ \citenamefont
  {Yanagida}}]{Nakayama:2018yvj}%
  \BibitemOpen
  \bibfield  {author} {\bibinfo {author} {\bibfnamefont {K.}~\bibnamefont
  {Nakayama}}, \bibinfo {author} {\bibfnamefont {F.}~\bibnamefont {Takahashi}},
  \ and\ \bibinfo {author} {\bibfnamefont {T.~T.}\ \bibnamefont {Yanagida}},\
  }\href {\doibase 10.1016/j.physletb.2019.01.013} {\bibfield  {journal}
  {\bibinfo  {journal} {Phys. Lett.}\ }\textbf {\bibinfo {volume} {B790}},\
  \bibinfo {pages} {218} (\bibinfo {year} {2019})},\ \Eprint
  {http://arxiv.org/abs/1811.01755} {arXiv:1811.01755 [hep-ph]} \BibitemShut
  {NoStop}%
%%CITATION = ARXIV:1811.01755;%%
\bibitem [{\citenamefont {Poulin}\ \emph {et~al.}(2016)\citenamefont {Poulin},
  \citenamefont {Serpico},\ and\ \citenamefont {Lesgourgues}}]{Poulin:2016nat}%
  \BibitemOpen
  \bibfield  {author} {\bibinfo {author} {\bibfnamefont {V.}~\bibnamefont
  {Poulin}}, \bibinfo {author} {\bibfnamefont {P.~D.}\ \bibnamefont {Serpico}},
  \ and\ \bibinfo {author} {\bibfnamefont {J.}~\bibnamefont {Lesgourgues}},\
  }\href {\doibase 10.1088/1475-7516/2016/08/036} {\bibfield  {journal}
  {\bibinfo  {journal} {JCAP}\ }\textbf {\bibinfo {volume} {1608}},\ \bibinfo
  {pages} {036} (\bibinfo {year} {2016})},\ \Eprint
  {http://arxiv.org/abs/1606.02073} {arXiv:1606.02073 [astro-ph.CO]}
  \BibitemShut {NoStop}%
%%CITATION = ARXIV:1606.02073;%%
\bibitem [{\citenamefont {Xiao}\ \emph {et~al.}(2020)\citenamefont {Xiao},
  \citenamefont {Zhang}, \citenamefont {An}, \citenamefont {Feng},\ and\
  \citenamefont {Wang}}]{Xiao:2019ccl}%
  \BibitemOpen
  \bibfield  {author} {\bibinfo {author} {\bibfnamefont {L.}~\bibnamefont
  {Xiao}}, \bibinfo {author} {\bibfnamefont {L.}~\bibnamefont {Zhang}},
  \bibinfo {author} {\bibfnamefont {R.}~\bibnamefont {An}}, \bibinfo {author}
  {\bibfnamefont {C.}~\bibnamefont {Feng}}, \ and\ \bibinfo {author}
  {\bibfnamefont {B.}~\bibnamefont {Wang}},\ }\href {\doibase
  10.1088/1475-7516/2020/01/045} {\bibfield  {journal} {\bibinfo  {journal}
  {JCAP}\ }\textbf {\bibinfo {volume} {2001}},\ \bibinfo {pages} {045}
  (\bibinfo {year} {2020})},\ \Eprint {http://arxiv.org/abs/1908.02668}
  {arXiv:1908.02668 [astro-ph.CO]} \BibitemShut {NoStop}%
%%CITATION = ARXIV:1908.02668;%%
\bibitem [{\citenamefont {Enqvist}\ \emph {et~al.}(2015)\citenamefont
  {Enqvist}, \citenamefont {Nadathur}, \citenamefont {Sekiguchi},\ and\
  \citenamefont {Takahashi}}]{Enqvist:2015ara}%
  \BibitemOpen
  \bibfield  {author} {\bibinfo {author} {\bibfnamefont {K.}~\bibnamefont
  {Enqvist}}, \bibinfo {author} {\bibfnamefont {S.}~\bibnamefont {Nadathur}},
  \bibinfo {author} {\bibfnamefont {T.}~\bibnamefont {Sekiguchi}}, \ and\
  \bibinfo {author} {\bibfnamefont {T.}~\bibnamefont {Takahashi}},\ }\href
  {\doibase 10.1088/1475-7516/2015/09/067} {\bibfield  {journal} {\bibinfo
  {journal} {JCAP}\ }\textbf {\bibinfo {volume} {1509}},\ \bibinfo {pages}
  {067} (\bibinfo {year} {2015})},\ \Eprint {http://arxiv.org/abs/1505.05511}
  {arXiv:1505.05511 [astro-ph.CO]} \BibitemShut {NoStop}%
%%CITATION = ARXIV:1505.05511;%%
\bibitem [{\citenamefont {Enqvist}\ \emph {et~al.}(2019)\citenamefont
  {Enqvist}, \citenamefont {Nadathur}, \citenamefont {Sekiguchi},\ and\
  \citenamefont {Takahashi}}]{Enqvist:2019tsa}%
  \BibitemOpen
  \bibfield  {author} {\bibinfo {author} {\bibfnamefont {K.}~\bibnamefont
  {Enqvist}}, \bibinfo {author} {\bibfnamefont {S.}~\bibnamefont {Nadathur}},
  \bibinfo {author} {\bibfnamefont {T.}~\bibnamefont {Sekiguchi}}, \ and\
  \bibinfo {author} {\bibfnamefont {T.}~\bibnamefont {Takahashi}},\ }\href@noop
  {} {\  (\bibinfo {year} {2019})},\ \Eprint {http://arxiv.org/abs/1906.09112}
  {arXiv:1906.09112 [astro-ph.CO]} \BibitemShut {NoStop}%
%%CITATION = ARXIV:1906.09112;%%
\bibitem [{\citenamefont {Tremaine}\ and\ \citenamefont
  {Gunn}(1979)}]{Tremaine:1979we}%
  \BibitemOpen
  \bibfield  {author} {\bibinfo {author} {\bibfnamefont {S.}~\bibnamefont
  {Tremaine}}\ and\ \bibinfo {author} {\bibfnamefont {J.~E.}\ \bibnamefont
  {Gunn}},\ }\href {\doibase 10.1103/PhysRevLett.42.407} {\bibfield  {journal}
  {\bibinfo  {journal} {Phys. Rev. Lett.}\ }\textbf {\bibinfo {volume} {42}},\
  \bibinfo {pages} {407} (\bibinfo {year} {1979})},\ \bibinfo {note}
  {[,66(1979)]}\BibitemShut {NoStop}%
%%CITATION = PRLTA,42,407;%%
\bibitem [{\citenamefont {Boyarsky}\ \emph {et~al.}(2009)\citenamefont
  {Boyarsky}, \citenamefont {Ruchayskiy},\ and\ \citenamefont
  {Iakubovskyi}}]{Boyarsky:2008ju}%
  \BibitemOpen
  \bibfield  {author} {\bibinfo {author} {\bibfnamefont {A.}~\bibnamefont
  {Boyarsky}}, \bibinfo {author} {\bibfnamefont {O.}~\bibnamefont
  {Ruchayskiy}}, \ and\ \bibinfo {author} {\bibfnamefont {D.}~\bibnamefont
  {Iakubovskyi}},\ }\href {\doibase 10.1088/1475-7516/2009/03/005} {\bibfield
  {journal} {\bibinfo  {journal} {JCAP}\ }\textbf {\bibinfo {volume} {0903}},\
  \bibinfo {pages} {005} (\bibinfo {year} {2009})},\ \Eprint
  {http://arxiv.org/abs/0808.3902} {arXiv:0808.3902 [hep-ph]} \BibitemShut
  {NoStop}%
%%CITATION = ARXIV:0808.3902;%%
\bibitem [{\citenamefont {Di~Paolo}\ \emph {et~al.}(2018)\citenamefont
  {Di~Paolo}, \citenamefont {Nesti},\ and\ \citenamefont
  {Villante}}]{DiPaolo:2017geq}%
  \BibitemOpen
  \bibfield  {author} {\bibinfo {author} {\bibfnamefont {C.}~\bibnamefont
  {Di~Paolo}}, \bibinfo {author} {\bibfnamefont {F.}~\bibnamefont {Nesti}}, \
  and\ \bibinfo {author} {\bibfnamefont {F.~L.}\ \bibnamefont {Villante}},\
  }\href {\doibase 10.1093/mnras/sty091} {\bibfield  {journal} {\bibinfo
  {journal} {Mon. Not. Roy. Astron. Soc.}\ }\textbf {\bibinfo {volume} {475}},\
  \bibinfo {pages} {5385} (\bibinfo {year} {2018})},\ \Eprint
  {http://arxiv.org/abs/1704.06644} {arXiv:1704.06644 [astro-ph.GA]}
  \BibitemShut {NoStop}%
%%CITATION = ARXIV:1704.06644;%%
\bibitem [{\citenamefont {Nakayama}\ \emph {et~al.}(2014)\citenamefont
  {Nakayama}, \citenamefont {Takahashi},\ and\ \citenamefont
  {Yanagida}}]{Nakayama:2013nya}%
  \BibitemOpen
  \bibfield  {author} {\bibinfo {author} {\bibfnamefont {K.}~\bibnamefont
  {Nakayama}}, \bibinfo {author} {\bibfnamefont {F.}~\bibnamefont {Takahashi}},
  \ and\ \bibinfo {author} {\bibfnamefont {T.~T.}\ \bibnamefont {Yanagida}},\
  }\href {\doibase 10.1016/j.physletb.2014.01.022} {\bibfield  {journal}
  {\bibinfo  {journal} {Phys. Lett.}\ }\textbf {\bibinfo {volume} {B730}},\
  \bibinfo {pages} {24} (\bibinfo {year} {2014})},\ \Eprint
  {http://arxiv.org/abs/1311.4253} {arXiv:1311.4253 [hep-ph]} \BibitemShut
  {NoStop}%
%%CITATION = ARXIV:1311.4253;%%
\bibitem [{\citenamefont {Harigaya}\ \emph {et~al.}(2013)\citenamefont
  {Harigaya}, \citenamefont {Kawasaki},\ and\ \citenamefont
  {Yanagida}}]{Harigaya:2012hn}%
  \BibitemOpen
  \bibfield  {author} {\bibinfo {author} {\bibfnamefont {K.}~\bibnamefont
  {Harigaya}}, \bibinfo {author} {\bibfnamefont {M.}~\bibnamefont {Kawasaki}},
  \ and\ \bibinfo {author} {\bibfnamefont {T.~T.}\ \bibnamefont {Yanagida}},\
  }\href {\doibase 10.1016/j.physletb.2013.01.002} {\bibfield  {journal}
  {\bibinfo  {journal} {Phys. Lett.}\ }\textbf {\bibinfo {volume} {B719}},\
  \bibinfo {pages} {126} (\bibinfo {year} {2013})},\ \Eprint
  {http://arxiv.org/abs/1211.1770} {arXiv:1211.1770 [hep-ph]} \BibitemShut
  {NoStop}%
%%CITATION = ARXIV:1211.1770;%%
\bibitem [{\citenamefont {Mangano}\ and\ \citenamefont
  {Serpico}(2011)}]{Mangano_2011}%
  \BibitemOpen
  \bibfield  {author} {\bibinfo {author} {\bibfnamefont {G.}~\bibnamefont
  {Mangano}}\ and\ \bibinfo {author} {\bibfnamefont {P.~D.}\ \bibnamefont
  {Serpico}},\ }\href {\doibase 10.1016/j.physletb.2011.05.075} {\bibfield
  {journal} {\bibinfo  {journal} {Physics Letters B}\ }\textbf {\bibinfo
  {volume} {701}},\ \bibinfo {pages} {296} (\bibinfo {year}
  {2011})}\BibitemShut {NoStop}%
\bibitem [{\citenamefont {Borzumati}\ \emph {et~al.}(2008)\citenamefont
  {Borzumati}, \citenamefont {Bringmann},\ and\ \citenamefont
  {Ullio}}]{Bringmann:2007ft}%
  \BibitemOpen
  \bibfield  {author} {\bibinfo {author} {\bibfnamefont {F.}~\bibnamefont
  {Borzumati}}, \bibinfo {author} {\bibfnamefont {T.}~\bibnamefont
  {Bringmann}}, \ and\ \bibinfo {author} {\bibfnamefont {P.}~\bibnamefont
  {Ullio}},\ }\href {\doibase 10.1103/PhysRevD.77.063514} {\bibfield  {journal}
  {\bibinfo  {journal} {Phys. Rev.}\ }\textbf {\bibinfo {volume} {D77}},\
  \bibinfo {pages} {063514} (\bibinfo {year} {2008})},\ \Eprint
  {http://arxiv.org/abs/hep-ph/0701007} {arXiv:hep-ph/0701007 [hep-ph]}
  \BibitemShut {NoStop}%
%%CITATION = HEP-PH/0701007;%%
\bibitem [{\citenamefont {Cembranos}\ \emph {et~al.}(2005)\citenamefont
  {Cembranos}, \citenamefont {Feng}, \citenamefont {Rajaraman},\ and\
  \citenamefont {Takayama}}]{Cembranos:2005us}%
  \BibitemOpen
  \bibfield  {author} {\bibinfo {author} {\bibfnamefont {J.~A.~R.}\
  \bibnamefont {Cembranos}}, \bibinfo {author} {\bibfnamefont {J.~L.}\
  \bibnamefont {Feng}}, \bibinfo {author} {\bibfnamefont {A.}~\bibnamefont
  {Rajaraman}}, \ and\ \bibinfo {author} {\bibfnamefont {F.}~\bibnamefont
  {Takayama}},\ }\href {\doibase 10.1103/PhysRevLett.95.181301} {\bibfield
  {journal} {\bibinfo  {journal} {Phys. Rev. Lett.}\ }\textbf {\bibinfo
  {volume} {95}},\ \bibinfo {pages} {181301} (\bibinfo {year} {2005})},\
  \Eprint {http://arxiv.org/abs/hep-ph/0507150} {arXiv:hep-ph/0507150 [hep-ph]}
  \BibitemShut {NoStop}%
%%CITATION = HEP-PH/0507150;%%
\bibitem [{\citenamefont {Colin}\ \emph {et~al.}(2000)\citenamefont {Colin},
  \citenamefont {Avila-Reese},\ and\ \citenamefont
  {Valenzuela}}]{Colin:2000dn}%
  \BibitemOpen
  \bibfield  {author} {\bibinfo {author} {\bibfnamefont {P.}~\bibnamefont
  {Colin}}, \bibinfo {author} {\bibfnamefont {V.}~\bibnamefont {Avila-Reese}},
  \ and\ \bibinfo {author} {\bibfnamefont {O.}~\bibnamefont {Valenzuela}},\
  }\href {\doibase 10.1086/317057} {\bibfield  {journal} {\bibinfo  {journal}
  {Astrophys. J.}\ }\textbf {\bibinfo {volume} {542}},\ \bibinfo {pages} {622}
  (\bibinfo {year} {2000})},\ \Eprint {http://arxiv.org/abs/astro-ph/0004115}
  {arXiv:astro-ph/0004115 [astro-ph]} \BibitemShut {NoStop}%
%%CITATION = ASTRO-PH/0004115;%%
\bibitem [{\citenamefont {Choi}\ \emph
  {et~al.}(2020{\natexlab{b}})\citenamefont {Choi}, \citenamefont {Suzuki},\
  and\ \citenamefont {Yanagida}}]{Choi:2020nan}%
  \BibitemOpen
  \bibfield  {author} {\bibinfo {author} {\bibfnamefont {G.}~\bibnamefont
  {Choi}}, \bibinfo {author} {\bibfnamefont {M.}~\bibnamefont {Suzuki}}, \ and\
  \bibinfo {author} {\bibfnamefont {T.}~\bibnamefont {Yanagida}},\ }\href@noop
  {} {\  (\bibinfo {year} {2020}{\natexlab{b}})},\ \Eprint
  {http://arxiv.org/abs/2004.07863} {arXiv:2004.07863 [hep-ph]} \BibitemShut
  {NoStop}%
\end{thebibliography}%

\end{document}